\pdfoutput=1
\RequirePackage{fix-cm}
\documentclass[smallextended]{svjour3}       
\smartqed  
\usepackage{graphicx}
\graphicspath{./figures}
\usepackage{amsmath,amssymb}
\usepackage{bm}
\usepackage[mathscr]{eucal}
\usepackage{booktabs}
\usepackage{siunitx}
\usepackage{multirow}

\usepackage[dvipsnames]{xcolor}
\usepackage{colortbl}

\usepackage[switch,displaymath,mathlines,modulo]{lineno}

\def\be{\begin{equation}}
\def\ee{\end{equation}}
\def\bea{\begin{eqnarray}}
\def\eea{\end{eqnarray}}

\def\({\left(}
\def\){\right)}

\def\x{\mathbf{x}}
\def\z{\mathbf{z}}
\def\y{\mathbf{y}}

\def\f{\mathbf{f}}

\def\g{\mathbf{g}}

\def\H{\mathcal{H}}

\def\bF{\mathbf{F}}
\def\bG{\mathbf{G}}

\def\bR{\mathbf{R}}

\def\bM{\mathbf{M}}

\def\epsi{{\bm \epsilon}}

\def\Re{{\mathbb R}}

\def\b1{{\mathbf 1}}

 \journalname{Journal of Statistical Physics}

\begin{document}

\title{On temporal scale separation in coupled data assimilation with the ensemble Kalman filter
}


\author{Maxime Tondeur$^{1}$ \and
        Alberto Carrassi$^{2,3,4}$ \and
        Stephane Vannitsem$^{5}$ \and
        Marc Bocquet$^{6}$ 
}


\institute{ \at
              (1) Paris, France 
           \and
              \at
              (2) Nansen Environmental and Remote Sensing Center, Bergen, Norway \\
	      (3) Mathematical Institute, Utrecht University, Utrecht, Holland\\	
	      (4) Dept of Meteorology, University of Reading and National Centre for Earth Observations, United Kingdom 
              \email{n.a.carrassi@reading.ac.uk}           
           \and
              \at
	      (5) Royal Meteorological Institute of Belgium, Brussels, Belgium	
           \and
              \at
	      (6) CEREA, joint laboratory \'Ecole des Ponts ParisTech and EDF R\&D, Universit\'e Paris-Est, Champs-sur-Marne, France
}

\date{Received: date / Accepted: date}

\maketitle


\begin{abstract}

Data assimilation for systems possessing many scales of motions is a substantial methodological and technological challenge.
Systems with these features are found in many areas of computational physics and are becoming common thanks to increased computational power allowing to resolve finer scales and to couple together several sub-components. Coupled data assimilation (CDA) distinctively appears as a main concern in numerical weather and climate prediction with major efforts put forward by meteo services worldwide. The core issue is the scale separation acting as a barrier that hampers the propagation of the information across model components ({\it e.g.} ocean and atmosphere).

We provide a brief survey of CDA, and then focus on CDA using the ensemble Kalman filter (EnKF), a widely used Monte Carlo Gaussian method. Our goal is to elucidate the mechanisms behind information propagation across model components. 
We consider first a coupled system of equations with temporal scale difference, and deduce that: (i) cross components effects are strong from the slow to the fast scale, 
but, (ii) intra-component effects are much stronger in the fast scale. While observing the slow scale is desirable and benefits the fast, the latter must be observed with high frequency otherwise the error will grow up to affect the slow scale.
 
Numerical experiments are performed using the atmosphere-ocean model, MAOOAM. Six configurations are considered, differing for the strength of the atmosphere-ocean coupling and/or the number of model modes. The performance of the EnKF depends on the model configuration, {\it i.e.} on its dynamical features. 
A comprehensive dynamical characterisation of the model configurations is provided by examining the Lyapunov spectrum, Kolmogorov entropy and Kaplan-Yorke attractor dimension. We also compute the covariant Lyapunov vectors and use them to explain how model instabilities act on different model's modes according to the coupling strength. 

The experiments  confirm the importance of observing the fast scale, but show also that, despite its slow temporal scale, frequent observations in the ocean are beneficial.
The relation between the ensemble size, $N$, and the unstable subspace dimension, $n_0$, has been studied. Results largely ratify what known for uncoupled system: the condition $N\ge n_0$ is necessary for the EnKF to work satisfactorily. Nevertheless the quasi-degeneracy of the Lyapunov spectrum of MAOOAM, with many near-zero exponents, is potentially the cause of the smooth gradual reduction of the analysis error observed for some model configurations, even when $N>n_0$.
Future prospects for the EnKF in the context of coupled ocean-atmosphere systems are finally discussed.

\end{abstract}

\section{Introduction}
\label{sec:intro}

Data assimilation (DA) is the term used to refer to a broad family of conceptual, mathematical and numerical methods performing the combination of model solutions with data from observing devices of a system. A popular terrain of applications of DA, and one that distinguishes DA from more general classes of optimisation or filtering methods, is its widespread use in the context of chaotic dynamics. The primary goal is to get an estimate of the system's state that is more accurate than the ones given by the model and data independently (see {\it e.g.}, \cite{asch2016data}). Data assimilation posed its root in the geosciences, particularly meteorology, but its use is becoming widespread across other areas of the geosciences and beyond \cite{carrassi2018data}. Examples include, but are not limited to, the attribution of climate change \cite{Hannart-et-al-2016}, neuro- ({\it e.g.} \cite{kadakia2016nonlinear,moye2018data}) and life- sciences \cite{hutt2019data} or traffic management \cite{palatella2013nonlinear}. 

This work is about DA for systems possessing a wide range of spatial and temporal scales, in particular coupled dynamical systems, in which the typical temporal scales of the system's components are different and generally not overlapping. This situation is common in physical science and it arises when modelling a continuum system in high resolution. Similarly when the system is modelled by coupling together different sub-systems each one spanning its own band of spatio-temporal scales. 
Notable examples are the climate models that couple together the different components of the Earth system. But it is nowadays present in other domains, including computational biology, neuroscience or economy, wherever the today’s enhanced computational power allows to explicitly couple several sub-systems at an almost constantly increasing resolution.

Together with the increase in models’ resolution, computational geosciences in the last decade has also seen the increase of prediction lengths beyond the meteorological time horizon of two weeks.
Seasonal-to-decadal (s2d) forecasts, a time horizon bearing enormous societal relevance, are possible because predictable signals arise from the interaction between the fast ({\it e.g.}, the atmosphere) and the slow ({\it e.g.}, the ocean, the land surface or the cryosphere) varying components of the system \cite{doblas2013seasonal}.
In the range between weather and s2d predictions stands the sub-seasonal to seasonal (s2s) time range,  which corresponds to predictions from two weeks to a season \cite{brunet2010collaboration}.
Sub-seasonal to seasonal predictions have motivated the ongoing transition toward the so-called ``seamless'' weather/climate prediction, in which the same fully coupled climate model is used to predict from minutes to months (see {\it e.g.}, \cite{palmer2008toward,brunet2015seamless}).

Coupled data assimilation (CDA) is needed to enhance the predictive skill of phenomena connected to the air-sea exchange like hurricanes or coastal weather, or in s2d predictions where climate conditions are triggered by coupled processes such as ENSO. The development of efficient CDA methods has been identified as crucial already in the assessment report of the $5{\rm th}$ Intergovernmental Panel on Climate Change, and several research institutions, are involved in studying and developing CDA (see, {\it e.g.}, \cite{penny2017coupled}).

From a DA perspective, the main issue is that the scale separation renders it extremely difficult to carry out the uncertainty quantification necessary to propagate consistently the information from the observations in one component throughout the full system.
If the scale separation is not very large, one can still rely upon standard, uncoupled DA that operates on each component independently, and then use the full coupled model to propagate information between successive observations, an approach known as {\it weakly CDA} (wCDA). Although in wCDA the effect of the coupling manifests indirectly via the model forward integration, the cross-component physical correlations (if any) are not exploited in the analysis update. Such a procedure is thus prone to produce imbalances, and a fully CDA (usually referred to as {\it strongly CDA}, sCDA) is required.

We focus on sequential DA methods, such as the ensemble Kalman filter (EnKF, \cite{evensen2009}), with the aim of elucidating the nature of the problem. 
We will develop our discussion in relation to the geosciences, where the field has been pushed forward. However none of the results nor of the conclusions are restricted to that context exclusively. 

\subsection{Coupled data assimilation in the geosciences: brief survey}
\label{sec:intro2}

We provide here a succinct survey of CDA efforts in the geosciences that is functional to our discussion. Recent reviews of CDA can be found in \cite{penny2017coupled,penny2017coupledWMO,Penny-2019}. In particular \cite{Penny-2019} provides a detailed comparison of different DA methods using the same low dimensional coupled model used here (see Sect.~\ref{sec:MAOOAM})
 
Early attempts include a Kalman filter type approach used to assimilate sparse data in a system with various spatio-temporal scales \cite{harlim2010filtering}, and the EnKF in a two-scale low dimensional system \cite{ballabrera2009data}. 
A modification of the 4-dimensional variational assimilation (4DVar; see {\it e.g.} \cite{carrassi2018data} its Section~3.2 and references therein) for coupled dynamics is given in \cite{lorenc20074d}, although the approach presented a number of practical issues making difficult its application in an operational scenario. 

First wCDA reanalyses have been obtained at the USA National Centre for Environmental Prediction \cite{saha2010ncep} and at the Japanese Agency for Marine-Earth Science and Technology \cite{sugiura2008development}, with global coupled models using 3DVar and 4DVar respectively. 
At the European Centre for Medium Range Weather Forecast, wCDA-like is performed with the incremental 4DVar in an innovative way. 
While all other terms are ({\it i.e.} background error covariance and observation operator) are kept uncoupled, the full coupled model is used in the outer loop of the minimisation, resulting in an implicit coupling that manifest within the assimilation window \cite{laloyaux2016coupled}. The method has been used to produce the reanalyses CERA-20C \cite{laloyaux2018cera} for the entire $20{\rm th}$ century, and CERA-SAT \cite{schepers2018cera} that include satellite data. A comparison between the explicit ({\it i.e.} complete sCDA) and the implicit coupling in the incremental 4DVar has shown that for long assimilation window the latter produces accurate analysis, but the explicit coupling is preferable for short assimilation windows. The transition from a reanalysis to real time prediction is currently under study \cite{browne2019weakly}.

The EnKF in a wCDA setting has been successfully used to assimilate ocean data and initialise s2d predictions with the Norwegian Earth System Model (NorESM, \cite{counillon2014seasonal}).
Weakly CDA using the EnKF (in particular the Ensemble Adjustment one)  has been performed in \cite{zhang2007system} to constrain independently atmosphere and ocean at the Geophysical Fluid Dynamics Laboratory (GFDL).

The authors of \cite{lu2015strongly} proposed a sCDA approach in which the observed ocean-atmosphere correlation asymmetry is exploited explicitly when performing the coupled analysis. 
The maximum correlation occurred when the atmosphere leads the ocean by about the decorrelation time of the atmosphere.
The method is referred to as Leading Averaged Coupled Covariance (LACC) and the cross atmosphere-ocean covariance are constructed by using the leading ({\it i.e.} one decorrelation time ahead) forecasts and observations and the current ocean state.      
Using the local ensemble transform Kalman filter (LETKF, \cite{hunt2007}), the authors of \cite{sluka2016assimilating} improved over wCDA using only atmospheric observations in a coupled atmosphere–ocean model. 
Strongly coupled EnKF was implemented in \cite{tardif2014coupled} to recover the Atlantic meridional overturning circulation (AMOC) with simulated observations in a low-order coupled atmosphere-ocean model and later with averaged data of the atmosphere from a millennial-scale simulation of a comprehensive coupled atmosphere–ocean climate model \cite{tardif2015coupled}.
One of the first cases of an operational, EnKF-based, sCDA for a coupled ocean and sea ice model, is the Norwegian TOPAZ system \cite{sakov2012topaz4}.

The Maximum Likelihood Ensemble Kalman filter (MLFE, \cite{zupanski2005}) has also been successfully used in a number of sCDA applications \cite{vzupanski2017data}. These include land-atmosphere coupling \cite{suzuki2017case}, aerosol-atmosphere coupling \cite{acp-2019-2}, as well as chemistry-atmosphere coupling \cite{park2015structure}.  
Different 4DVar CDA approaches are discussed in \cite{smith2015exploring} using an idealised single-column atmosphere–ocean model, the estimation of the cross error covariances for use in CDA with 4DVar is studied in \cite{smith2017estimating}, while strategies to mitigate the sampling error in CDA have been described in \cite{smith2018treating}. 

\subsection{Outline}

An heuristic explanation of the impact of the temporal scale separation on CDA is provide in Sect.~\ref{sec:intro3}. The numerical atmosphere-ocean model MAOOAM is introduced in Sect.~\ref{sec:MAOOAM} together with a detailed analysis of its stability properties in connection to the atmosphere-ocean coupling strength. Definitions and significance of the Lyapunov exponents and vectors used for the stability analysis are recalled in the Appendix. Numerical experiments using an EnKF are given in Sect.~\ref{sec:DA}, followed by the conclusions in Sect.~\ref{sec:concl}.

\section{The nature of the problem}
\label{sec:intro3}

This section aims at illustrating key dynamical aspects of DA in coupled systems with time scale separation. We will intentionally set ourselves in a very idealised framework thus that the discussion that follows has only a general qualitative scope. With that in mind, our goal is to highlight: (i) which scale is more important to be observed, and, (ii) why it is desirable to allow observations from one component to impact the other.

Let us consider two coupled, deterministic and autonomous, ordinary differential equations (ODE)
as a prototype for a multiscale dynamical system

\be
\label{eq:MODEL}
\begin{aligned}
\frac{{\rm d}\x}{{\rm d}t} & = \epsilon\f(\x,\z), \\
\frac{{\rm d}\z}{{\rm d}t} & = \g(\x,\z),
\end{aligned}
\ee
with $\x\in\Re^{m_x}$, $\z\in\Re^{m_z}$, $\f:\Re^{m_x+m_z}\mapsto\Re^{m_x}$, $\g:\Re^{m_x+m_z}\mapsto\Re^{m_z}$.
The processes $\f$ and $\g$ are assumed to have the same time-scale, thus that their temporal scale difference is ``artificially'' fully accounted for by the constant, $\epsilon\ll1$, making the variable $\x$ slower than $\z$.
The time $t$ is adimensionalized with respect to the typical time scale of the fast variables. 
We furthermore assume that $\f$ and $\g$ have similar magnitude, are both bounded from above as $\mathcal{O}(1)$, and the characteristic spatial scales of $\x$ and $\z$ are similar. We recall the unrealistic characters of the above hypotheses. In particular the latter one is done here in order to simplify the treatment thus focusing on the effect of the timescale difference exclusively: in realistic coupled atmosphere-ocean models, atmosphere and ocean do have different spatial scales.

Within a time interval $t_k-t_{k-1}=\mathcal{O}(1)$ the slow scale changes such as $\mathcal{O}(\x_k)= \x_{k-1}+\mathcal{O}(\epsilon)$, while the fast scale as $\mathcal{O}(\z_k)= \z_{k-1}+\mathcal{O}(1)$.
Let us suppose to have observations of both scales, $\y^{\rm x}\in\Re^{\rm d_x}$ and $\y^{\rm z}\in\Re^{\rm d_z}$, for the slow and fast scale, respectively. 
Data from each scale are collected with different frequencies, proportional to their respective time scale, so that observations are more frequent for the fast than for the slow scale. 

In order for the observations of the slow scale system's component to monitor its variability, the observational interval has to be of order $\Delta t^{\rm x}=\mathcal{O}(\epsilon^{-1})$. The fast scale observational interval has to be shorter than the slow scale one, $\Delta t^{\rm z}\le\Delta t^{\rm x}$, and we stipulate for convenience that $\Delta t^{\rm x}=K\Delta t^{\rm z}$, with $K\in\mathbb{N}$, meaning that every $\Delta t^{\rm x}$ both scales are simultaneously observed.
Note that when $\Delta t^{\rm z}=\mathcal{O}(1)$ the solution of the slow system, $\x(t)$, can be considered approximately constant in the interval $t\in[t_k,t_k+\Delta t^{\rm z}]$.

The model defined by Eq.~\eqref{eq:MODEL} is used to assimilate recursively data $\y^{\rm z}$ every $\Delta t^{\rm z}$, and data $\y=(\y^{\rm x}, \y^{\rm z})^{\rm T}$ whenever $t_k$ is a multiple of $\Delta t^{\rm x}$. 
The linearised error evolution between two subsequent analyses reads

\be 
\label{eq:LinMODEL}
 \begin{bmatrix}
   \Delta\x_k^{\rm f} \\
   \Delta\z_k^{\rm f} 
  \end{bmatrix}
\approx 
 \begin{bmatrix}
   \epsilon\bF \\
    \bG
 \end{bmatrix}
 \begin{bmatrix}
   \Delta\x_{k-1}^{\rm a} \\ 
   \Delta\z_{k-1}^{\rm a}
 \end{bmatrix} 
=
  \begin{bmatrix}
    \epsilon\bF_{\x} & \epsilon\bF_{\z} \\
    \bG_{\x} & \bG_{\z}
  \end{bmatrix}
  \begin{bmatrix}
   \Delta\x_{k-1}^{\rm a} \\ 
   \Delta\z_{k-1}^{\rm a}
 \end{bmatrix}
\ee

where $\Delta\x_k$ and $\Delta\z_k$ are the errors in the slow and fast variables, respectively, while the super-scripts ``${\rm f}$'' and ``${\rm a}$'' stand for forecast and analysis. 
The terms $\bF_{\x}$ and $\bF_{\z}$ are components of the tangent linear model of $\f$, in particular its linearisation with respect to $\x$ and $\z$; the same applies to $\g$. 

We first consider how the error in one component impacts the other. This is regulated by the cross terms in Eq.~\eqref{eq:LinMODEL}, namely $\bF_{\z}\Delta\z_{k-1}^{\rm a}$ for the $Fast$-to-$Slow$ dependence, and $\bG_{\x}\Delta\x_{k-1}^{\rm a}$ for the $Slow$-to-$Fast$. Their amplitude measures the degree of sensitivity of one component to the other, and depend on the type and strength of the coupling.
For instance, atmosphere-ocean coupling, is usually described via two distinct, yet dependent, processes. A mechanical transfer of kinetic energy from the atmospheric wind to the ocean surface, that tends to slow down atmospheric wind and to enhance ocean waves, and a thermal coupling in which heat is transferred from the warmer to the colder model component. These processes would be, at the first order, encoded in the terms $\bF_{\z}$ and $\bG_{\x}$, and their relative dominance reflected in their amplitudes. 


Let us suppose that at the arbitrary analysis time, $t_{k-1}$, the analysis error on both components, $\Delta\x_{k-1}^{\rm a}$ and $\Delta\z_{k-1}^{\rm a}$, is of $\mathcal{O}(1)$. Using the linearised Eq.~\eqref{eq:LinMODEL} we can describe the first order error dynamics within the assimilation interval $\Delta t^{\rm z}$. Let us insert the error order in Eq.~\eqref{eq:LinMODEL} and take the norm of both sides
\be
\label{eq:LinMODELord}
\begin{aligned}
\mathcal{O}(\Delta\x_k^{\rm f}) & \approx \epsilon[\|\bF_{\x} + \bF_{\z}\|]\mathcal{O}(1) \le \epsilon[\|\bF_{\x}\| + \|\bF_{\z}\|]\mathcal{O}(1), \\    
\mathcal{O}(\Delta\z_k^{\rm f}) & \approx [\|\bG_{\x} + \bG_{\z}\|]\mathcal{O}(1) \le [\|\bG_{\x}\| + \|\bG_{\z}\|]\mathcal{O}(1).     
\end{aligned}
\ee

Given that the analysis error is of $\mathcal{O}(1)$, one desires (at the best) the forecast error bound to be also $\mathcal{O}(1)$. By substituting $\mathcal{O}(\Delta\x_k^{\rm f})=\mathcal{O}(1)$ and $\mathcal{O}(\Delta\z_k^{\rm f})=\mathcal{O}(1)$ in Eq.~\eqref{eq:LinMODELord}, we get the following bounds for the amplitude of the tangent linear model ({\it i.e.} the first order model sensitivity) of the slow component
\be
\label{eq:BoundSlow}
\begin{aligned}
\mathcal{O}(\|\bF_{\x}\|) \le \mathcal{O}(\epsilon^{-1}) & \quad Slow\mapsto Slow \ {\rm sensitivity}, \\
\mathcal{O}(\|\bF_{\z}\|) \le \mathcal{O}(\epsilon^{-1}) & \quad Fast\mapsto Slow \ {\rm sensitivity}, 
\end{aligned}
\ee    
and for the fast component
\be
\label{eq:BoundFast}
\begin{aligned}
\mathcal{O}(\|\bG_{\x}\|) \le \mathcal{O}(1) & \quad Slow\mapsto Fast \ {\rm sensitivity}, \\
\mathcal{O}(\|\bG_{\z}\|) \le \mathcal{O}(1) & \quad Fast\mapsto Fast \ {\rm sensitivity}. 
\end{aligned}
\ee    
From Eq.~\eqref{eq:BoundSlow}, we see that the slow scale sensitivities can be as large as $\mathcal{O}(\epsilon^{-1})$. This means that an $\mathcal{O}(1)$ error in any of the scales will not (in general, $\mathcal{O}(\|\bF_{\x}\|),\mathcal{O}(\|\bF_{\z}\|)>\mathcal{O}(\epsilon^{-1})$) cause a larger order error in the slow scale forecast. In particular, the second inequality in \eqref{eq:BoundSlow} indicates that the $Fast$-to-$Slow$ scale effect is generally little, and the forecast error will not grow over $\mathcal{O}(1)$, within $\Delta t^{\rm z}=\mathcal{O}(1)$.   

The reduced $Fast$-to-$Slow$ effect is also explained by recalling that within the interval $\Delta t^{\rm z}$ the slow scale is almost constant and it largely ``feels'' the fast one via its smoothed averaged signal, with a time variability of the same order the slow scale. This mechanism is often adduced to explain the somehow little $Fast$-to-$Slow$ effect observed in coupled DA experiments with more realistic atmosphere-ocean coupled models. For instance \cite{tardif2015coupled} performed coupled DA with the EnKF in a comprehensive coupled atmosphere–ocean climate model showing that atmospheric observations alone, albeit frequent, do not suffice to properly recover the slowly evolving Atlantic meridional overturning circulation (AMOC), and that, in the absence of ocean data, the use of time-averaged atmospheric measurements was able to successfully track the AMOC (see also \cite{penny2017coupled} for a review of recent coupled DA operational efforts). Note however that, in those cases the fast and slow components do not generally have the same amplitude nor the same spatial scale, as we have hypothesised here.        

The sensitivity bounds on the fast scale, Eq.~\eqref{eq:BoundFast}, are smaller: an $\mathcal{O}(1)$ internal or $Slow$-to-$Fast$ sensitivity is enough to cause an $\mathcal{O}(1)$ forecast error growth. In particular, and as opposed to the $Fast$-to-$Slow$ case, the first of the inequalities \eqref{eq:BoundFast}, indicates the larger impact of the slow scale on the fast one, again in line with the aforementioned works by \cite{tardif2015coupled,penny2017coupled}.

Nevertheless, it is the second inequalities in \eqref{eq:BoundFast} that sets the highest challenge: it implies that the fast scale analysis error must be kept within $\mathcal{O}(1)$ otherwise a ``locally'' large $\|\bG_{\z}\|$, beyond $\mathcal{O}(1)$, will lead the forecast error to grow over $\mathcal{O}(1)$. 
The only way to achieve this is by directly observing the fast scale and, via coupled DA, allowing the slow scale measurements to update the fast scale. Whenever the fast scale is left unobserved, the error in the scale will generally grow over $\mathcal{O}(1)$ within $\Delta t^{\rm z}=\mathcal{O}(1)$ and, through the cross component sensitivity $\|\bF_{\z}\|$, will inevitably impact the slow scale too.

In conclusion, while the ideal situation is to have data on both scales, those on the fast one are particularly important. They are needed to keep error in the fast scale to a low level, thus preventing the growth of error in the slow scale via the crossing term. Slow scale observations are beneficial and desirable too. They are instrumental to keep error in the slow scale to small levels; they are, however, less capable to contain the growth of the fast scale errors. 
It is finally worth stressing again the very ideal character of the above conclusions and of Sect.~\ref{sec:intro3} at large. The full picture in a real system can be far more complicated. For instance the relative roles of the atmosphere and ocean in real system is observed to be very different in the Tropics and in mid-latitudes \cite{bach2019local}. 


  

\section{A coupled atmosphere-ocean model: MAOOAM}
\label{sec:MAOOAM}

\subsection{Generalities}
\label{sec:MAOOAMa}

In our experiments we shall use the coupled atmosphere-ocean numerical model MAOOAM \cite{DeCruz_et_al_2016}, which is an instructive low-order model for coupled dynamics. MAOOAM is composed of a two-layer quasi-geostrophic (QG) atmosphere that is coupled, both thermally and mechanically, to a QG shallow-water ocean layer in the $\beta$-plane approximation.
The model solves for the vorticity and the temperature in both media, and is written in spectral Fourier modes, whose full number can be adjusted to the desired resolution. 

In our applications we set the total number of Fourier modes alternatively to $m=36$, $52$, or $56$. Linear and nonlinear terms in the Fourier expansion are projected onto the phase subspace spanned by the selected modes, using an appropriate scalar product. The model and its properties are described in \cite{DeCruz_et_al_2016,VANNITSEM201571}. 

MAOOAM develops baroclinic instability: the solar forcing induces a horizontal North-South temperature gradient in the atmosphere, which in turns maintains the vertical gradient of the wind. This is possible because the atmosphere possesses two vertical layers. 
The wind gradient then creates a shear force which is responsible for eddies at the interface of the two layers; they are the cause of instability in the model. Concurrently, the ocean transports the heat to counteract the initial gradient of temperature.

The model is numerically integrated with a time-step of approximately $16$ minutes. 

\subsection{Selected model configurations}
\label{sec:MAOOAMb}

We consider three model setups with dimension $m=36$, $m=52$ and $m=56$. 
In the case, $m=36$ (the most widely used in previous studies with MAOOAM) the modes, {\it i.e.} the model's state vector components, are distributed between atmosphere and ocean as follows: the first $10$ are associated to the atmospheric barotropic streamfunction, followed by $10$ modes for the atmospheric temperature, $8$ for the ocean streamfunction, and $8$ for the ocean temperature. In the configuration $m=52$, $16$ modes ($8$ for both streamfunction and temperature respectively) are added to the ocean. Finally, for $m=56$, $4$ additional atmospheric modes ($2$ for both barotropic streamfunction and temperature) are added to the atmosphere. 

For each of these three model's dimensions, we consider two atmosphere-ocean couplings, hereafter referred to as $weak$ and $strong$, making a total of six model configurations: $36wk$, $52wk$, $56wk$, $36st$, $52st$ and $56st$. The coupling strength is varied by acting on the friction coefficients and the heat exchange between the two media, as described in Tab.~\ref{tab:1}; other key model parameters are included in  Tab.~\ref{tab:2}.

\begin{table}
\caption{Summary of the six MAOOAM configurations under consideration, with indication of the atmosphere-ocean coupling strength. The table reports the values of the key parameters, modulating the coupling's strength: the friction coefficients at the bottom of the atmosphere, $k$, between internal atmospheric layers, $kp$, and between atmosphere and ocean, $d$, as well as the heat exchange between the two media, $\lambda$, and the stationary solutions for the $0{\rm -th}$ order atmospheric and ocean temperature, $T^{\rm atm}_0$ and $T^{\rm ocn}_0$ (see \cite{vannitsem2015role} for a complete description of the model parameters and their role). The model dimension, for the atmosphere and ocean, $m^{\rm atm}$ and $m^{\rm ocn}$, respectively, is reported in the last two columns for each of the configurations. The total dimension is $m=m^{\rm atm}+m^{\rm ocn}$.}
\label{tab:1}
\begin{center}
\begin{tabular}{cccccccccc} \toprule
    & {$Coupling$} & {$k \ [adim] $} & {$kp \ [adim] $} & {$\lambda \ [\tfrac{W}{m^{2}K}] $} & {$d \ [s^{-1}] $} &  {$T^{{\rm atm}}_0 \ [K]$} & {$T^{{\rm ocn}}_0 \ [K]$} & {$m^{{\rm atm}}$} & {$m^{{\rm ocn}}$}  \\ \toprule
    {\bf 36wk} & \multirow{3}{*}{Weak} & \multirow{3}{*}{0.010} & \multirow{3}{*}{0.020} & \multirow{3}{*}{10} & \multirow{3}{*}{6$\times10^{-8}$}  & \multirow{3}{*}{289} &  \multirow{3}{*}{301} & 20 & 16   \\
                 {\bf 52wk} &	& & & &  & & & 20 & 32  \\ 
      		{\bf 56wk} &	& & &  & & & & 24 & 32   \\ \midrule
    {\bf 36st} & \multirow{3}{*}{Strong}  & \multirow{3}{*}{0.0145}   & \multirow{3}{*}{0.029} & \multirow{3}{*}{15.06} & \multirow{3}{*}{9$\times10^{-8}$} & \multirow{3}{*}{290.20} &  \multirow{3}{*}{299.35} & 20 & 16  \\
                 {\bf 52st} &       & & &  & & & & 20 & 32  \\
                 {\bf 56st} &   & & &  & & & &  14 & 32  \\ \bottomrule
\end{tabular}

\end{center}
\end{table}

\begin{table}
\caption{List of the remaining MAOOAM parameters having the same values in both coupling configurations. }
\label{tab:2}
\begin{center}
\begin{tabular}{ccl} \toprule
     {$C^{\rm atm} \ [\tfrac{W}{m^{2}}]$} & 310 &  Radiation input for the atmosphere  \\ \midrule
    {$C^{\rm ocn} \ [\tfrac{W}{m^{2}}]$} & 310 &  Net short wave radiation input for the ocean \\ \midrule
    {$\gamma^{\rm atm}  \ [\tfrac{J}{m^{2}K]}$} & $10^7$  &  Specific heat capacity of the atmosphere \\ \midrule
    {$\gamma^{\rm ocn}  \ [\tfrac{J}{m^{2}K]}$} & $6.6\times 10^8$  &  Specific heat capacity of the ocean \\ \midrule
    {$H  \ [m]$} & 165  &  Depth of the ocean layer \\ \bottomrule
\end{tabular}

\end{center}
\end{table}

\begin{figure}
\caption{Illustration of the MAOOAM attractor in the configurations $36wk$ (a) and $36st$ (b) for the three variables $(\psi^{{\rm ocn}}_2,\theta^{{\rm ocn}}_2,\psi^{{\rm atm}}_1)$. The model is integrated forward for $10^7$ days.}
\label{fig:1}
\begin{tabular}{cc}
 & \\
{\bf (a) - 36wk} & {\bf (b) - 36st} \\
\includegraphics[clip=true,width=0.49\columnwidth]{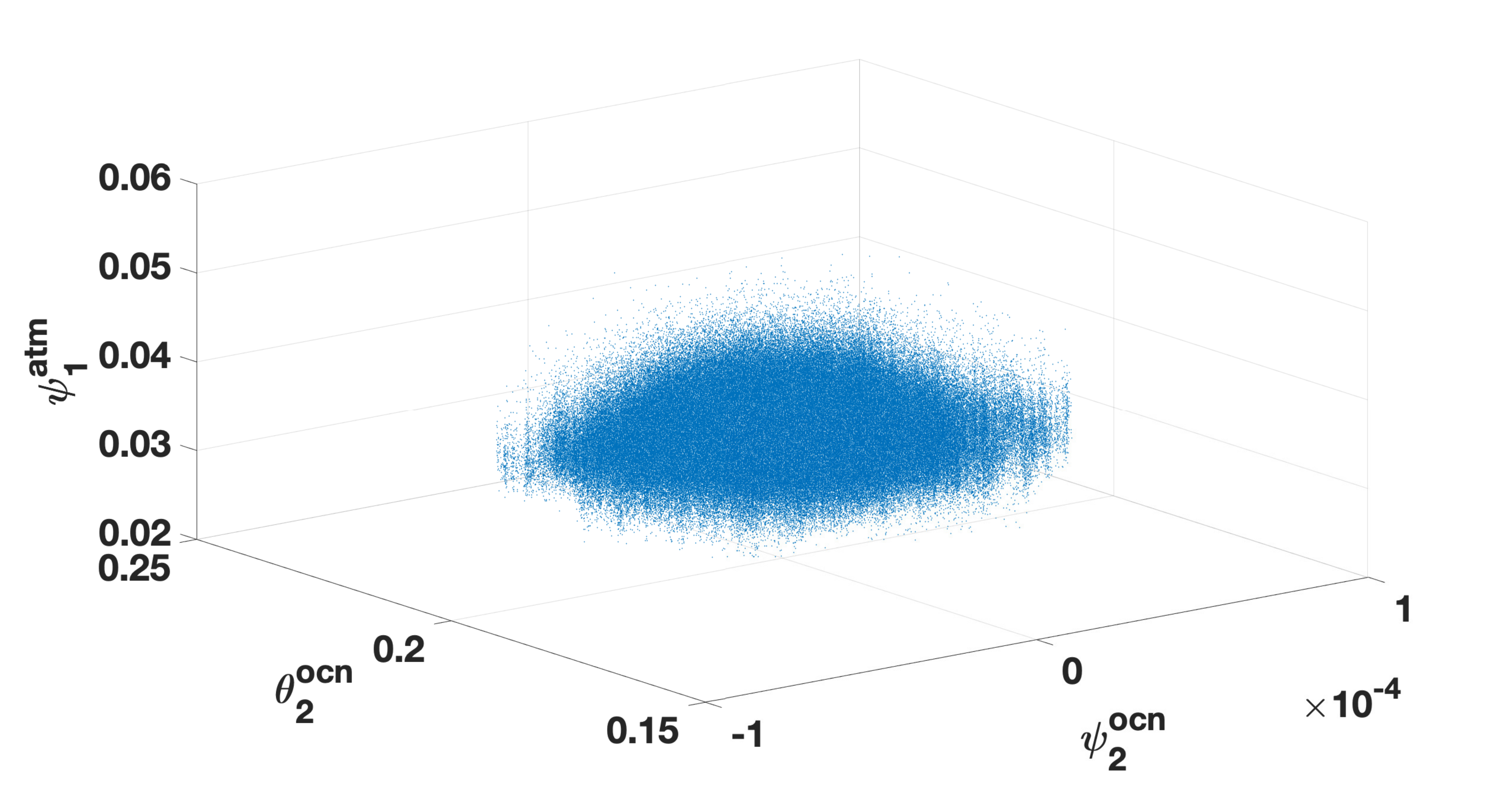} & 
\includegraphics[clip=true,width=0.49\columnwidth]{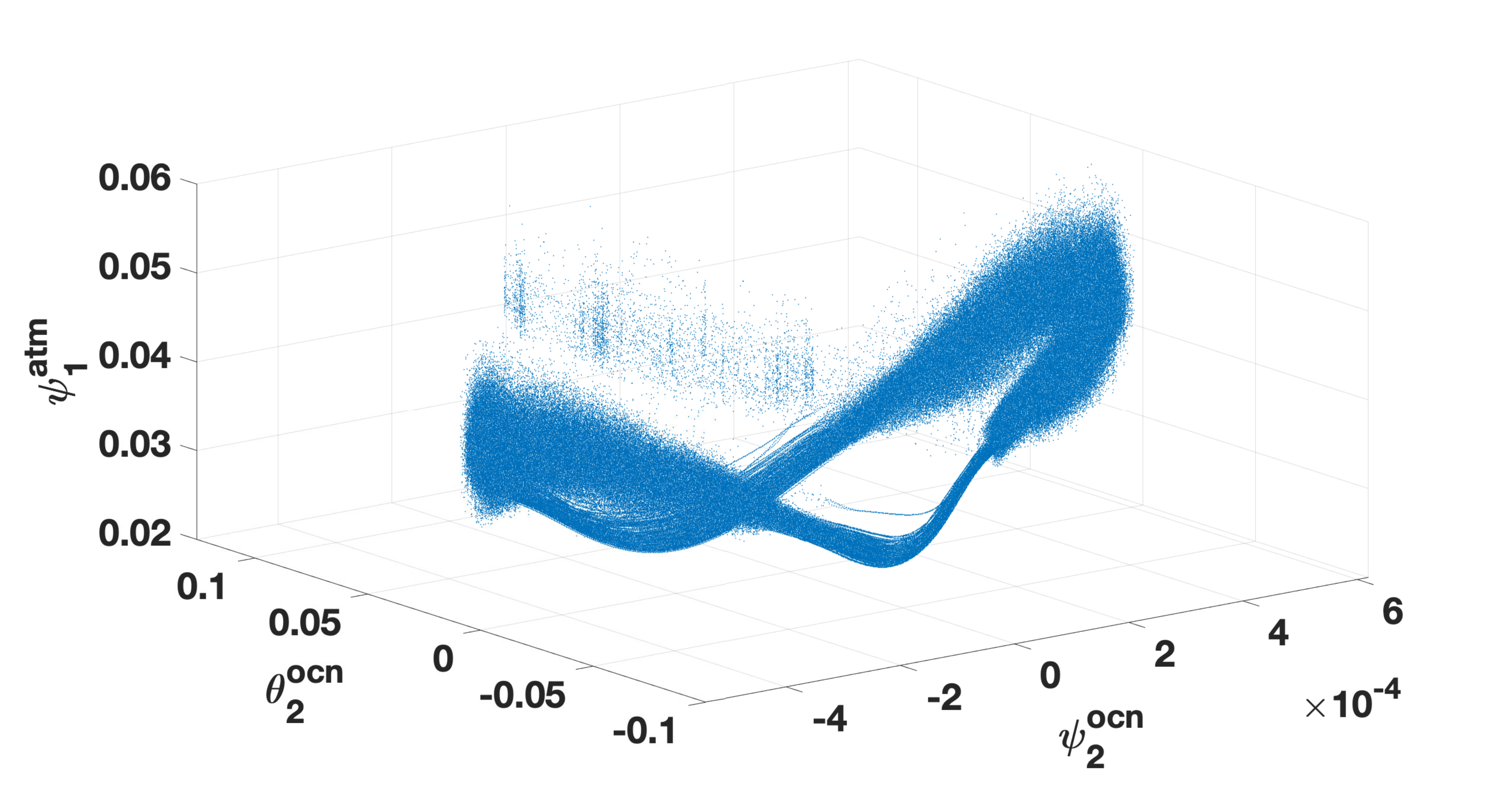}
\end{tabular}
\end{figure}

An illustration of the long term dynamical behaviour of configurations $36wk$ and $36st$ is given in Fig.~\ref{fig:1} (panels (a) and (b), respectively). Both panels show the trajectory solution of the model for $10^{7}$ days, projected onto the $3$-dimensional portion of the phase space spanned by three key modes  $(\psi^{{\rm ocn}}_2,\theta^{{\rm ocn}}_2,\psi^{{\rm atm}}_1)$, {\it i.e.} the second Fourier modes of the ocean streamfunction and temperature, and the first one of the atmospheric streamfunction; the importance of these three modes as representative of the model dynamics in the full phase space has been put forward in \cite{VANNITSEM201571}.

\begin{figure}
\begin{center}
\caption{Time series of the first component of the atmospheric streamfunction, $\psi^{{\rm atm}}_1$, for the model configuration 36st. The red and green points indicate the start of active and passive regimes respectively.}
\label{fig:2}
\includegraphics[clip=true,width=0.8\columnwidth]{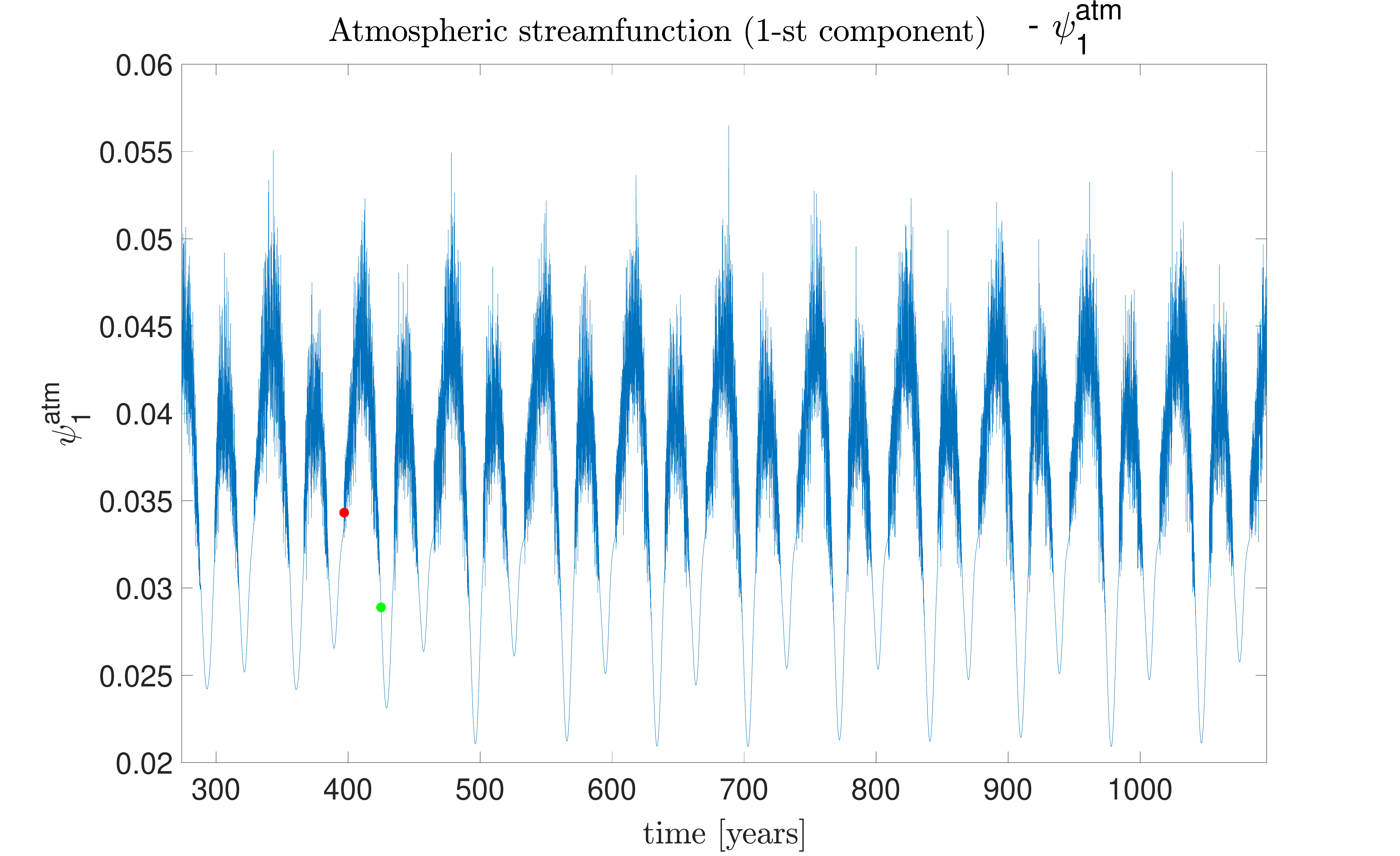}  
\end{center}
\end{figure}

The marked difference between the attractors' shapes (cf. the two panels of Fig.~\ref{fig:1}) is a manifestation of the different coupling strength. In the weakly coupled configuration, $36wk$ (panel (a)), the attractor has a sort of regular large scale shape (a spheroid) that is densely, albeit discontinuously due to chaos, filled by the trajectory as typical of an ergodic system. 
The attractor for the configuration $36st$ is still the one of a chaotic dynamics, yet it is organised now around an unstable periodic orbit around which the solution is wandering. This dynamics is accompanied by a succession of recurrences in regions of lower or higher values of the atmospheric streamfunction, $\psi^{{\rm atm}}_1$, with low and high variability respectively. We will hereafter refer to them as the ``passive'' and ``active'' regimes respectively. This different behaviour appears clear when looking at the time series of $\psi^{{\rm atm}}_1$ in Fig.~\ref{fig:2}; the red and green spots in the figure indicate the start of one active and one passive regime, respectively. Note furthermore that $\psi^{{\rm atm}}_1$ displays a low-frequency variability with a period of about $70$ years. This low-frequency variability is characterised by a slow motion along the attractor of the system leading for instance to the succession of peaks and minima in the streamfunction field of Fig.~\ref{fig:2}.

\begin{figure}
\caption{Correlation matrices in MAOOAM in configuration $36wk$; the axes display the system's state index. The correlation is computed every $10$ days, using the instantaneous values (a), and the atmospheric time-averaged values over the averaging period of $100$ days (b) and $1000$ days (c). Results are averaged over $10^5$ days. }
\label{fig:3}
\begin{tabular}{ccc}
 & \\
{\tiny {\bf (a) Instantaneous}} & {\tiny {\bf (b) $100$ days avgd atm}} & {\tiny {\bf (c) $1000$ days avgd atm}} \\
\includegraphics[clip=true,width=0.32\columnwidth]{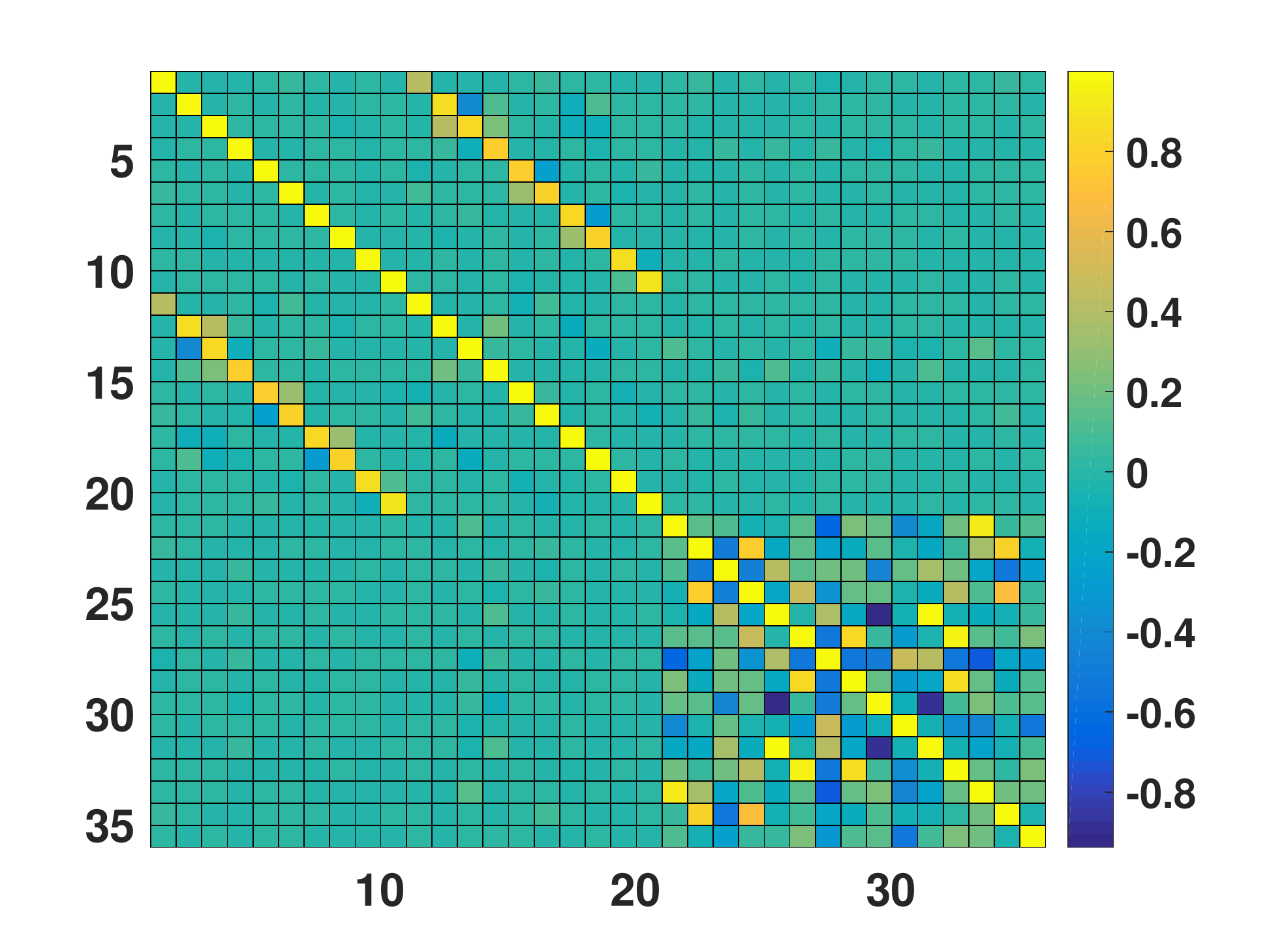} &
\includegraphics[clip=true,width=0.32\columnwidth]{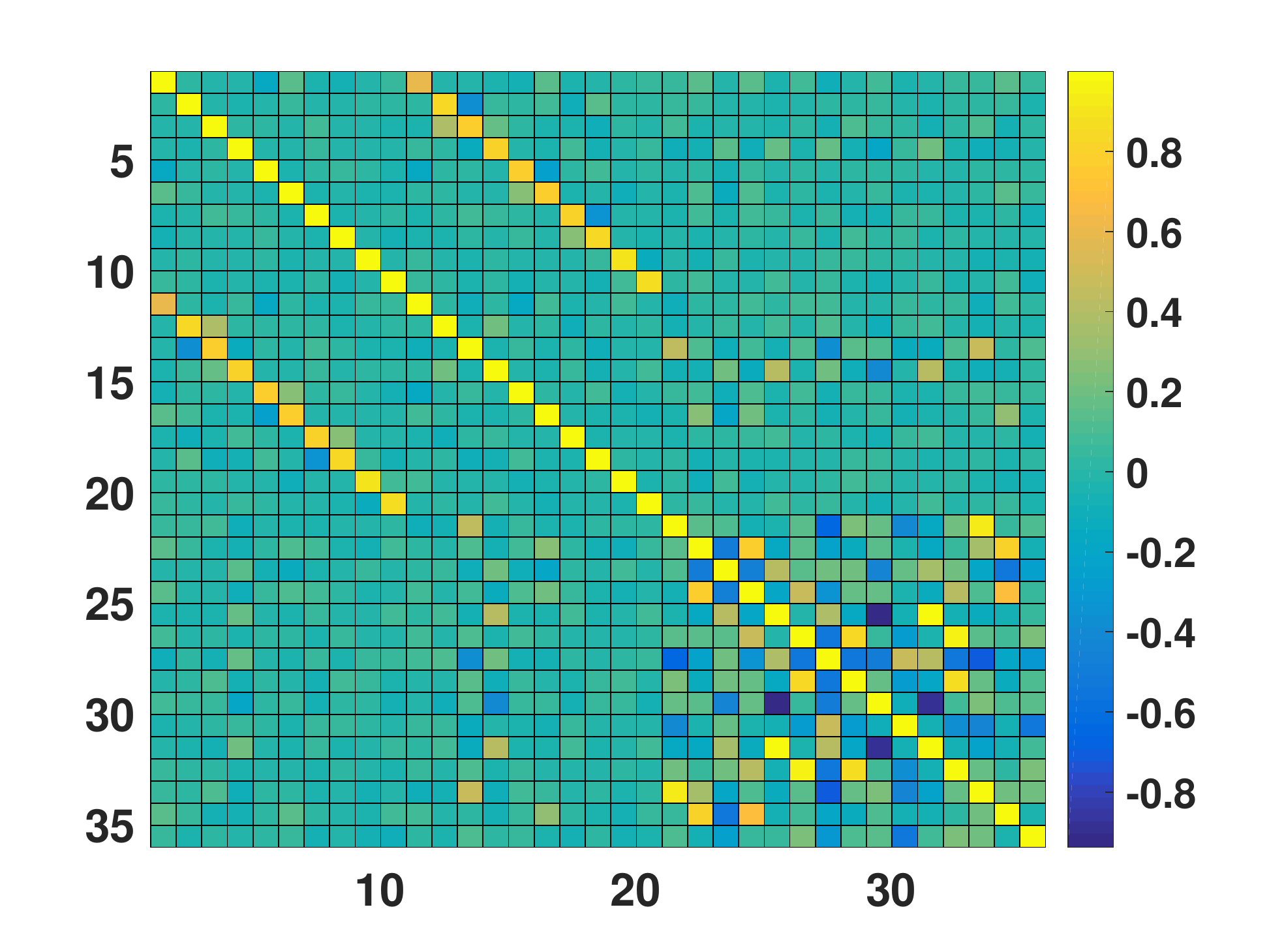} &
\includegraphics[clip=true,width=0.32\columnwidth]{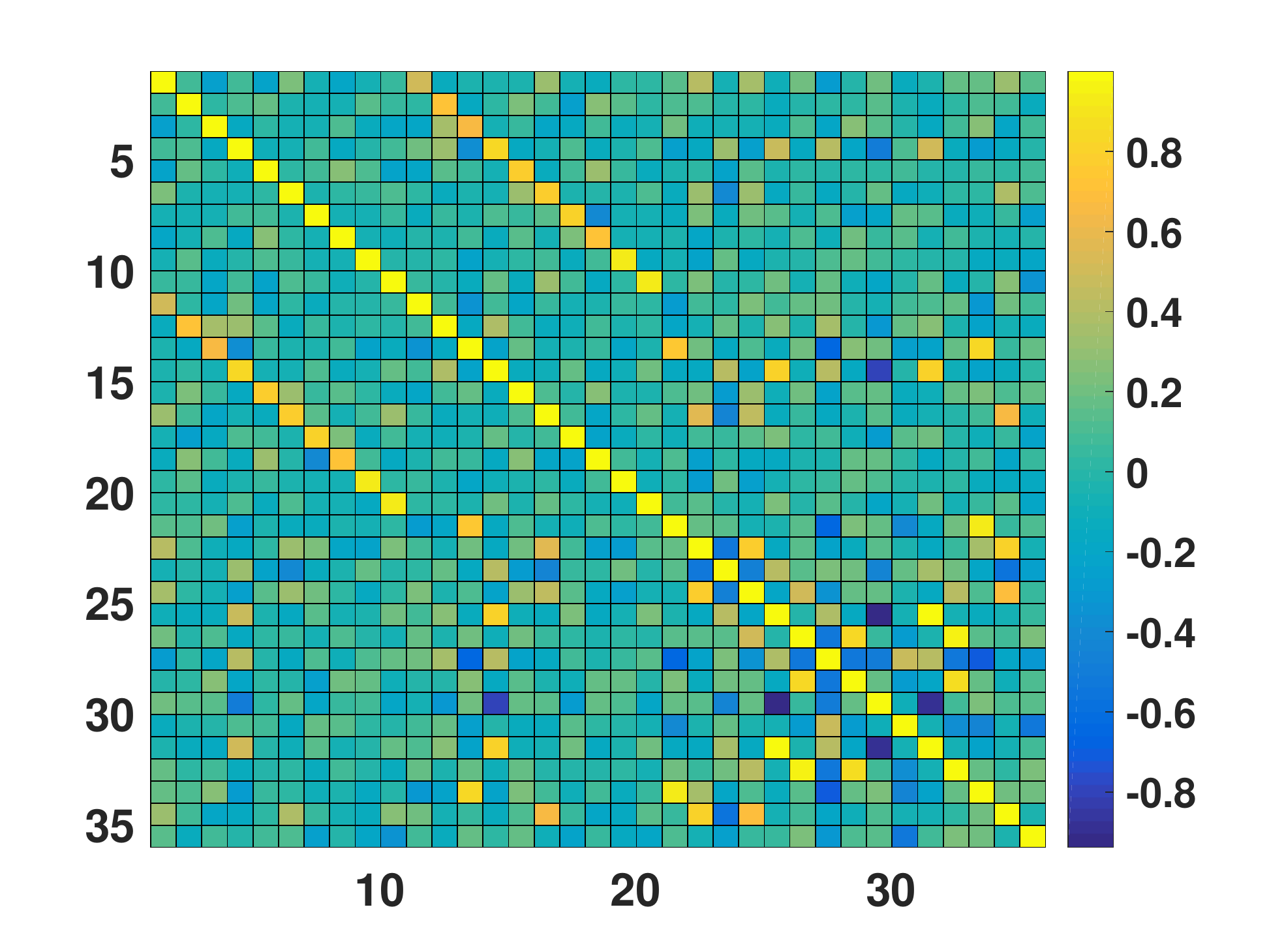}
\end{tabular}
\end{figure}

In order to estimate how the different variables in the model correlate to each other, and globally, how the atmosphere and ocean components are correlated, we compute the model auto-correlation every $10$ days and then averaged over $10^5$ days, for the configuration $36wk$. 
Results are shown in Fig.~\ref{fig:3} for three cases.
Besides the instantaneous values (panel (a)), we also compute the correlation between the ocean and the time-averaged atmosphere with averaging windows of $100$ days (panel (b)) and $1000$ days (panel (c)). 
  
Not surprisingly, when looking at the instantaneous values of the correlations (a), the self-components ({\it i.e.} the atmosphere-atmosphere and ocean-ocean) values are so much greater than the atmosphere-ocean correlation, that the latter values are almost invisible (yet they are not zero). It is interesting to note the well organised band-shape structure of the atmospheric correlation with a second maximum showing the correlation between atmospheric barotropic streamfunction and temperature, as opposed to the unstructured, yet very rich, pattern of the ocean auto-correlation. These are the correlations that would make possible in sCDA to update ocean/atmospheric variables by observing other ocean/atmospheric variables.

A noteworthy feature of Fig.~\ref{fig:3} is the substantial increase of the atmosphere-ocean cross correlation when the ocean is correlated with a time-averaged atmosphere (panels (b) and (c)). This cross correlation increases when the averaging window for the atmosphere is increased from $100$ to $1000$ days, and decreases further over $1000$ days (not shown). This behaviour naturally emerges as a consequence of the time-scale difference between ocean and atmosphere. It has been already put forward in previous studies (see {\it e.g.}  \cite{tardif2014coupled} and references therein), and is what has promoted the use of averaged observations in several early studies on coupled DA \cite{dirren2005toward,huntley2010assimilation,lu2015strongly}.%



\subsection{Stability analysis}
\label{sec:StabAn}

\begin{figure}
\begin{center}
\caption{Spectrum of Lyapunov exponents for the six MAOOAM configurations (top) and absolute values of the Lyapunov exponents (bottom) with log-scale in the $y-axis$.}
\label{fig:4}
\begin{tabular}{c}
\includegraphics[clip=true,width=0.8\columnwidth]{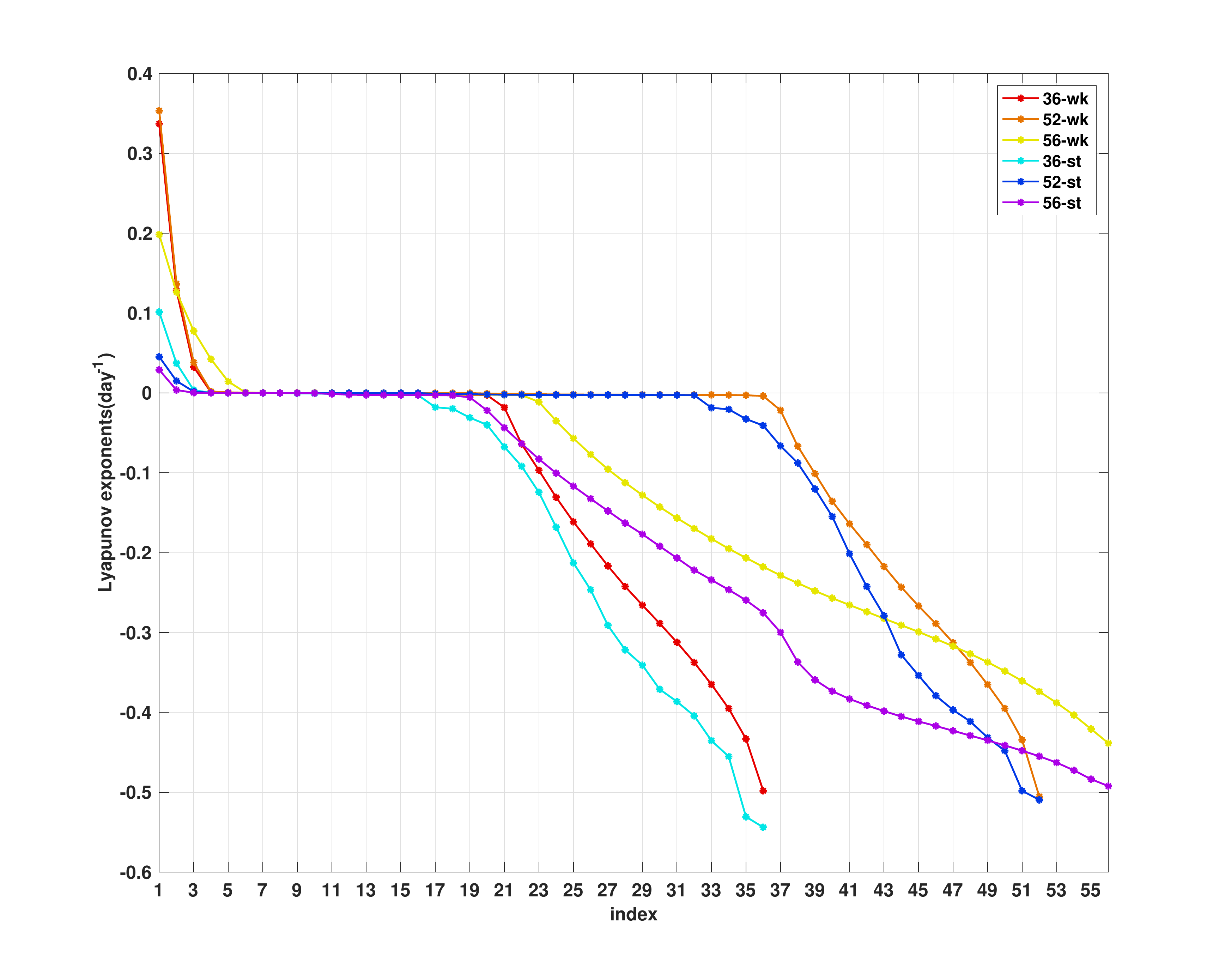} \\ 
\includegraphics[clip=true,width=0.8\columnwidth]{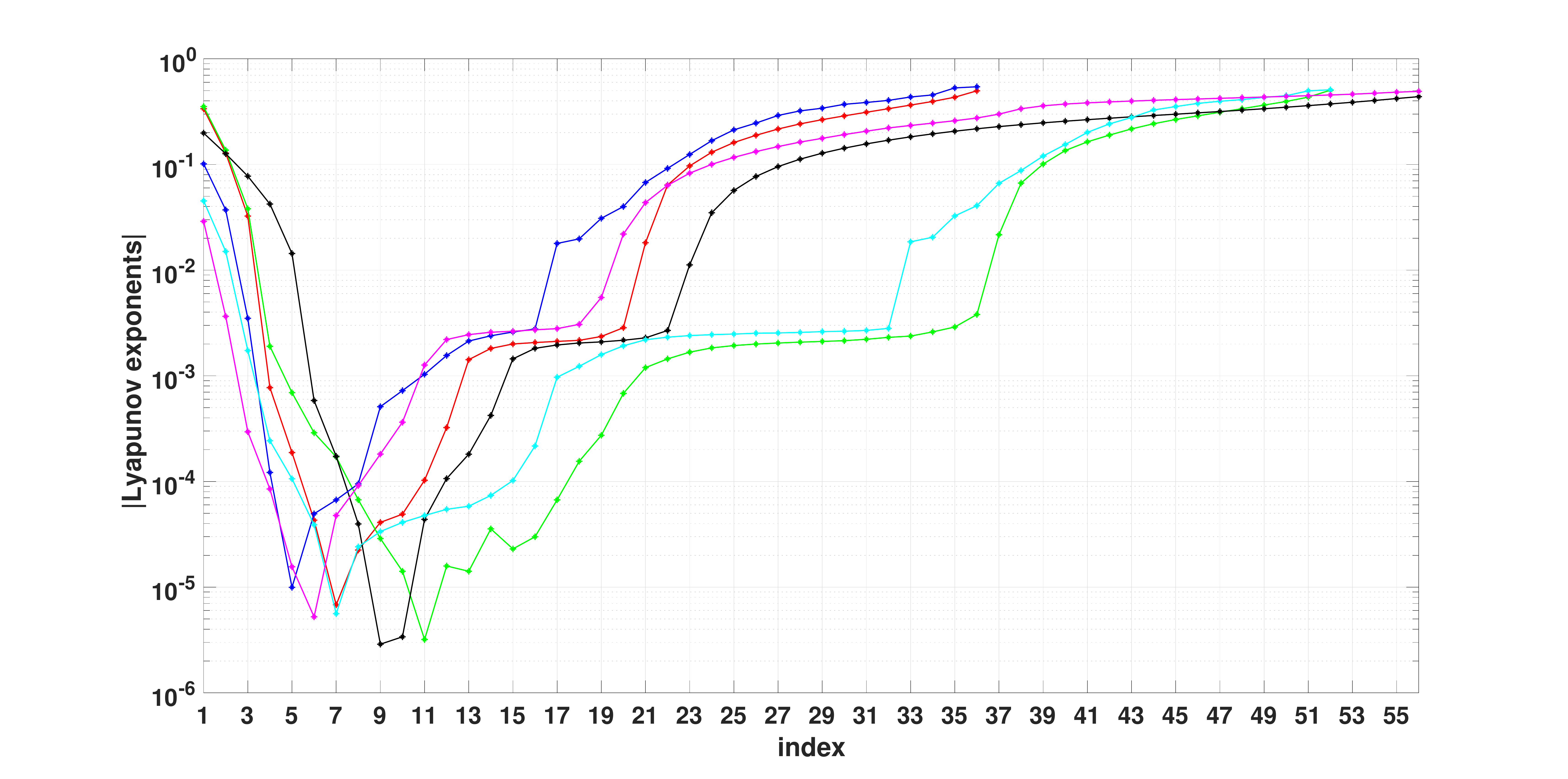}   
\end{tabular}
\end{center}
\end{figure}

We characterise the long-term dynamical behaviour of the six model configurations by studying their stability properties.
We compute their spectrum of Lyapunov exponents (LEs; see Appendix) and, based on them, the Kolmogorov entropy (KE; given by the sum of the positive LEs) and the Kaplan-Yorke attractor dimension (KY-dim) (see {\it e.g.}, \cite{ott2002chaos}). Results are reported in Tab.~\ref{tab:3}, while the spectrum's of the LEs for the six model configurations are shown in Fig.~\ref{fig:4}.
From Tab.~\ref{tab:3} we see that MAOOAM possesses a large number of almost neutral LEs. 
To better distinguish real neutral LEs (within numerical accuracy) from very little, albeit non-zero, ones, we split them in five categories: we will consider real neutral exponents those in the interval $\lambda_i\in[-10^{-5},\ 10^{-5}] $. The neighbouring ranges of ``near-neutral$^+$'' and ``near-neutral$^-$'' (see Tab.~\ref{tab:3}), encompass those exponents that, although not strictly zero, act almost as such.


\begin{table}
\caption{Summary of the stability analysis results for the six MAOOAM configurations. The numbers of LEs are counted as distributed in five ranges of values given in the first column. The last two rows report the values of KE and KY-dim respectively.}
\label{tab:3}
\begin{tabular}{lcccccc} \toprule
     {$Model\ Configuration$} & {\bf 36wk} & {\bf 52wk} & {\bf 56wk}  & {\bf 36st} & {\bf 52st} & {\bf 56st}  \\ \toprule
    {\bf \# Positive} \  $\lambda_i\in[10^{-2},\ 1]$ & 3 & 3 & 4 & 2  & 2 & 1   \\
    {\bf \# Near-neutral$^+$} \  $\lambda_i\in[10^{-5},\ 10^{-2}]$ & 3 & 7 & 3 & 2  & 4 & 4   \\
    {\bf \# Neutral} \  $\lambda_i\in[-10^{-5},\ 10^{-5}]$ & 1 & 1 & 2 & 1  & 1 & 1   \\
    {\bf \# Near-neutral$^-$} \  $\lambda_i\in[-10^{-2},\ -10^{-5}]$ & 13 & 25 & 12 & 11  & 25 & 13   \\
    {\bf \# Negative} \  $\lambda_i\in[-1,\ -10^{-2}]$ & 16 & 16 & 34 & 20  & 20 & 37   \\
    {\bf Kolmogorov entropy} \ & 0.498 & 0.528 & 0.459 & 0.139 & 0.060  & 0.029   \\
    {\bf Kaplan-Yorke dimension} \ & 25.06 & 41.03 & 28.42 & 20.29 & 33.35  & 19.32   \\ \bottomrule
\end{tabular}
\end{table}

As anticipated in Sect.~\ref{sec:MAOOAMa}, in the weakly coupled configurations the ocean slowing effect on the atmosphere is less effective, resulting in the model being more chaotic than in the corresponding strongly coupled configurations. By degree of chaos we mean the amplitude and number of the positive LEs. 
While both factors are merged in the definition of the KE, that in itself already represents a good measure of chaos, KE does not distinguish among the rate of growth along individual Lyapunov modes. All of the strong coupling configurations are characterised by smaller KEs, compared to the weakly ones.  

The addition of $20$ ocean modes from configuration $36wk/st$ to $52wk/st$ has almost no effect on the positive, nor on the negative, portions of the spectrum, but only on the neutral ones; as a result, the KE is only slightly different between $36wk/st$ and $52wk/st$. However, the KE is slightly larger in the $52wk$ case compared to $36wk$, and slightly smaller in the $52st$ compared to $36st$. This is because in the former case the amplitude of the positive LEs does not change much, while their number is larger for the configuration $52wk$. On the other hand, the larger number of positive LEs in the $52st$ over the $36st$ is counteracted by a reduction in their amplitudes.  

In both coupling strength cases, the transition from dimension $36$ to $52$ leads to almost doubling the number of the almost neutral LEs. These are a manifestation of, and are arisen by, the additional $20$ ocean modes. The role of the ocean modes as responsible for the neutral portion of the spectrum was already observed by \cite{vannitsem2016statistical}, where a broader analysis of the connection between physical variables and LEs in MAOOAM was presented. 
Note also that, although the increase of the almost neutral LEs does not change much the overall degree of instabilities (and therefore the intrinsic predictability of the configurations $36wk$ and $52wk$), it changes substantially the KY-dim, that is much larger in the $52wk$ case. In deterministic dynamics, the number of non-negative LEs, $n_0$, and the KY-dim are known to be directly proportional to the number of ensemble members that an ensemble Kalman filter (EnKF) needs to achieve satisfactory performance \cite{carrassi2009}, with $n_0$ being the minimum ensemble size required to avoid filter divergence \cite{bocquet2017four}. These findings have recently been explored for coupled dynamics by \cite{npg-27-51-2020}.   

The further dimensional increase from $52wk/st$ to $56wk/st$ causes a surprising, and difficult to interpret, change in the LEs spectrum. In both coupling cases, the number of positive (including small positive) LEs decreases, while that of negative LEs is doubled. 
Thus, the addition of the $4$ atmospheric modes is not increasing the degree of chaos as we might have expected based on the idea that atmosphere brings chaos, while ocean takes it away. The KE and the KY-dim are both smaller in the $56wk/st$ compared to $52wk/st$, implying that, despite the systems' state dimensions, {\it i.e.} the full phase-space, is larger, fewer ensemble members may be needed in the $56wk/st$ than in the $52wk/st$ configurations.     

\begin{figure}
\caption{Time average of the logarithm of the projections of the CLVs (y-axis) onto the model state vector components (index; x-axis) for configurations $36wk$ (a) and $36st$ (b).}
\label{fig:5}
\begin{tabular}{cc}
{\bf (a) - $36wk$} & {\bf (b) - $36st$} \\
\includegraphics[clip=true,width=0.45\columnwidth]{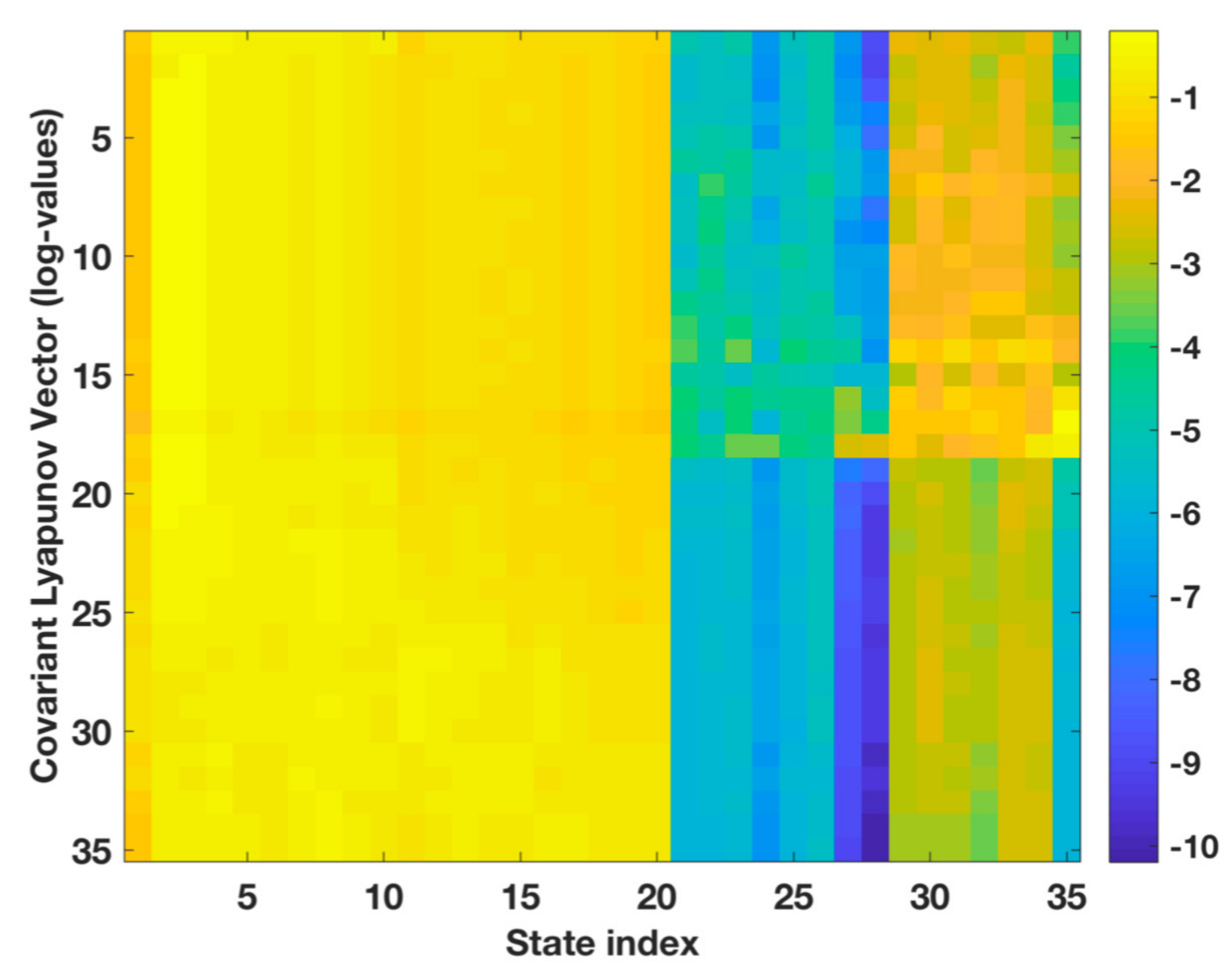} & 
\includegraphics[clip=true,width=0.44\columnwidth]{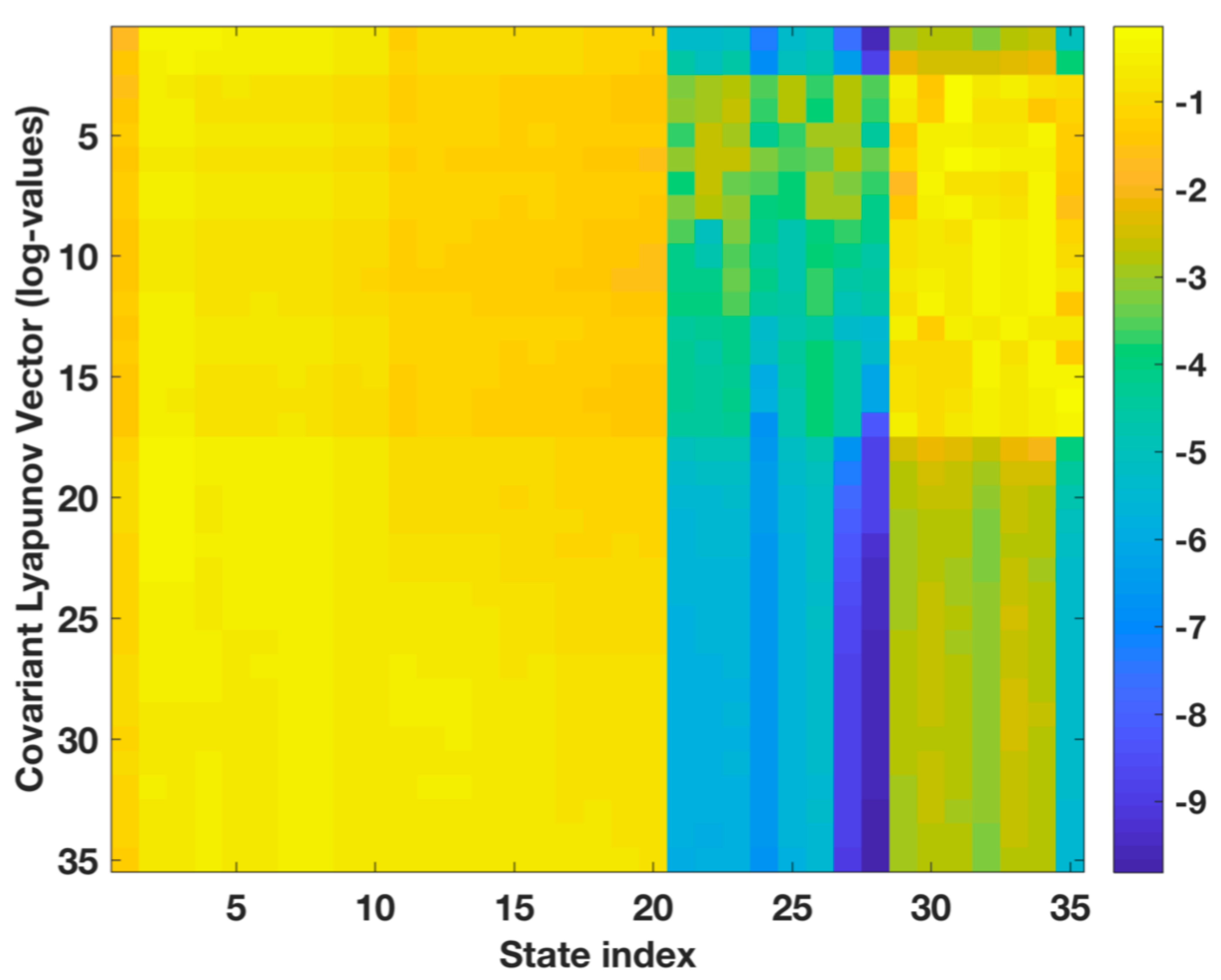}
\end{tabular}
\end{figure}

We conclude the section by studying the covariant Lyapunov vectors (CLVs; see Appendix) for the two smallest configurations, $36wk$ and $36st$. 
Similarly to the analysis reported in \cite{vannitsem2016statistical}, Fig.~\ref{fig:5} shows the CLVs amplitude projections (in log-scale) on the individual model's state vector components.   

Overall, the CLVs project largely on the atmospheric components ({\it i.e.} state index 1 to 20), but the oceanic temperature (state index $30$ to $36$) also presents significant projections. Some key CLVs associated with exponents close to 0 also display large averaged projections on the oceanic streamflow (state index 21 to 28). This demonstrates the coupling character of the instabilities that span across both atmosphere and ocean. In fact, when the coupling is increased (configuration $36st$, panel (b)) the relative amplitudes of the projections on the ocean components increase commensurately.

\section{Coupled ensemble Kalman filter with MAOOAM}
\label{sec:DA}

We present some illustrative numerical experiments using the ensemble Kalman filter (EnKF) with MAOOAM. The specific version of the EnKF adopted here is the {\it finite-size} EnKF (EnKF-N; \cite{bocquet2011,bocquet2015}). The EnKF-N is a deterministic EnKF with high accuracy in low-dimensional systems, and that incorporates the estimation of the inflation meant to counteract sampling errors, that would otherwise have had to be tuned. 
We do not report here the description of the EnKF-N; readers can find all details in \cite{bocquet2011,bocquet2015}. Note also that, to simplify the notation, we will hereafter systematically use the acronym EnKF to refer to the EnKF-N. This choice is also done to stress that the results that follow would be qualitatively the same for any deterministic formulation of the EnKF.    

The EnKF is used to perform sCDA and the results that follow refer to this case only. We have also performed experiments using wCDA and the results, not shown, indicate overall lower skills than sCDA. 

Experiments are performed with varying ensemble size, $N$, as well as atmosphere and ocean observational intervals, $\Delta t^{{\rm atm}}$ and $\Delta t^{{\rm ocn}}$. Simulated observations are sampled from a trajectory, solution of MAOOAM, that is taken to represent the truth with respect to which we compute the root-mean-square-error (RMSE), as in standard twin experiments. Observational error is simulated by adding Gaussian random noise, and the model-to-data relation reads:

\be
\label{eq:Hz}
\y_k = \y_k^{\rm atm} = \H^{\rm atm}(\x_k, \z_k) + \epsi^{{\rm atm}}_k \qquad {\rm when} \mod{(t_k,\Delta t^{\rm ocn})}\ne0, 
\ee
and 
\be
\y_k = 
\begin{bmatrix}
\y_k^{\rm ocn} \\
\y_k^{\rm atm} 
\end{bmatrix}
= 
\begin{bmatrix}
\H^{\rm ocn}(\x_k, \z_k) + \epsi^{\rm ocn}_k \\
\H^{\rm atm}(\x_k, \z_k) + \epsi^{\rm atm}_k
\end{bmatrix}
\qquad {\rm when} \mod{(t_k,\Delta t^{\rm ocn})}=0,
\ee
with $\H^{\rm ocn}:\Re^{m_{\rm ocn}+m_{\rm atm}}\mapsto\Re^{d_{\rm ocn}}$ and $\H^{\rm atm}:\Re^{m_{\rm ocn}+m_{\rm atm}}\mapsto\Re^{d_{\rm atm}}$ being the observational operators, $d_{\rm ocn}\le m_{\rm ocn}$ and $d_{\rm atm}\le m_{\rm atm}$, and $\Delta t^{\rm ocn}=K\Delta t^{\rm atm}$, $K\in\mathbb{N}$.
The observational errors in the two components, $\epsi^{\rm ocn}_k$ and $\epsi^{\rm atm}_k$, are assumed to be mutually independent and both unbiased and normally distributed with constant error covariances $\bR^{\rm ocn}\in\Re^{d_{\rm ocn}\times d_{\rm ocn}}$ and $\bR^{\rm atm}\in\Re^{d_{\rm atm}\times d_{\rm atm}}$, respectively. In the experiments the observational error is assumed to be spatially uncorrelated so that the matrices $\bR^{\rm ocn}$ and $\bR^{\rm atm}$ are diagonal, and the error standard deviation (the square-root of the diagonal entries of the matrices) to be equal to $1\%$ of the MAOOAM component-wise' natural variability, {\it i.e.} the long-term time averaged difference between uncorrelated states. MAOOAM variables are directly observed, implying that the observation operators $\H^{\rm ocn}$ and $\H^{\rm atm}$ are linear and expressed as matrices of appropriate dimension with only $1$ and $0$ as entries. The assumed ability to observe directly the model modes is an idealisation. In realistic scenarios one would have, at the best, point-wise measurements of physical quantities that are, usually nonlinearly, related to the model modes. This would introduce ``representation error'' (see {\it e.g.} \cite{janjic2018representation}) and degrade the performance of data assimilation as shown by \cite{Penny-2019} using MAOOAM.

\begin{figure}
\caption{
EnKF ensemble-based correlation matrices, with $N=15$ members, at time $t=1$ year of simulation. The axes display the system's state index.}
\label{fig:6}
\begin{tabular}{ccc}
 & \\
{\bf (a) 36wk} & {\bf (b) 36st active regime} & {\bf (c) 36st passive regime}\\
\includegraphics[clip=true,width=0.32\columnwidth]{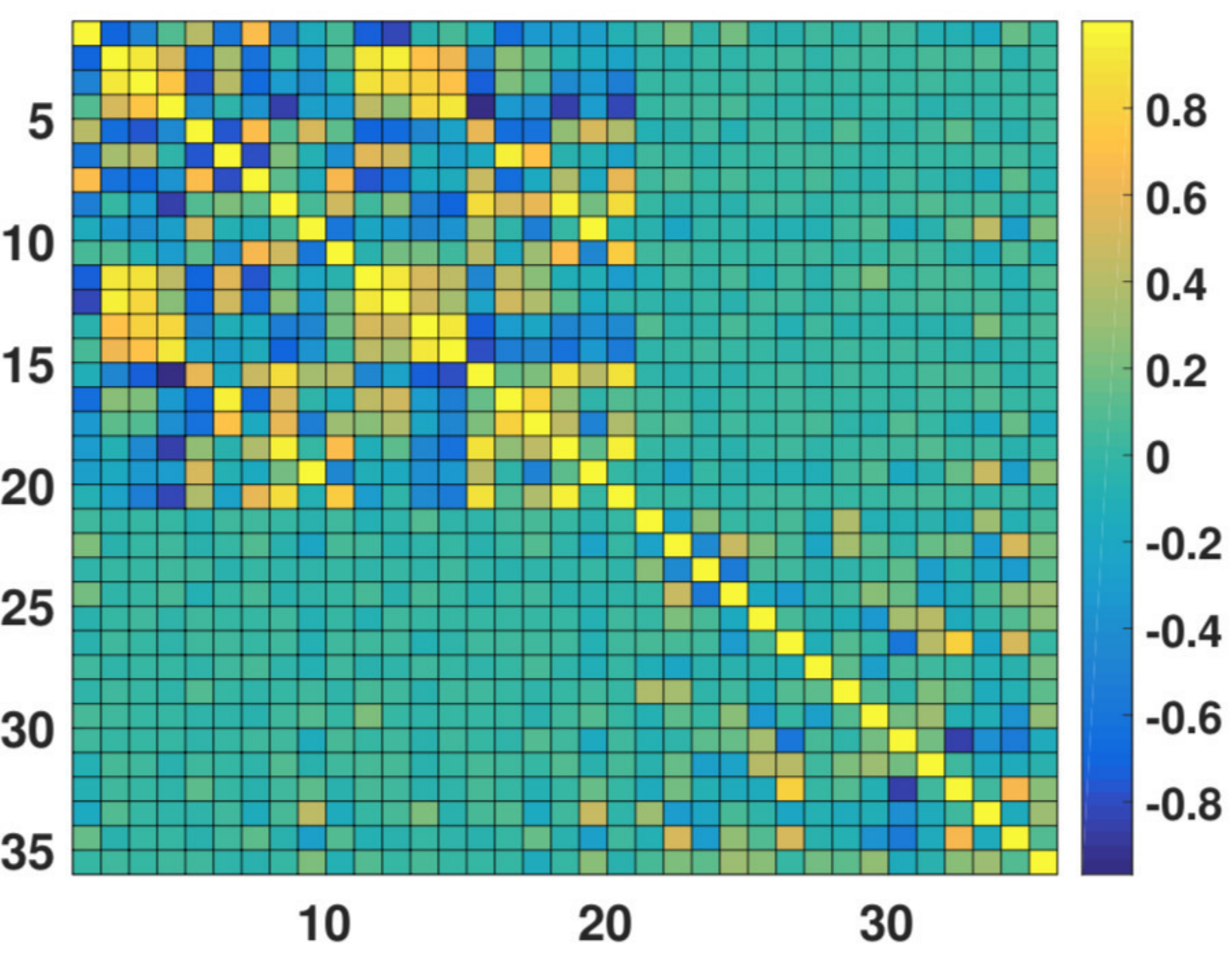} &
\includegraphics[clip=true,width=0.32\columnwidth]{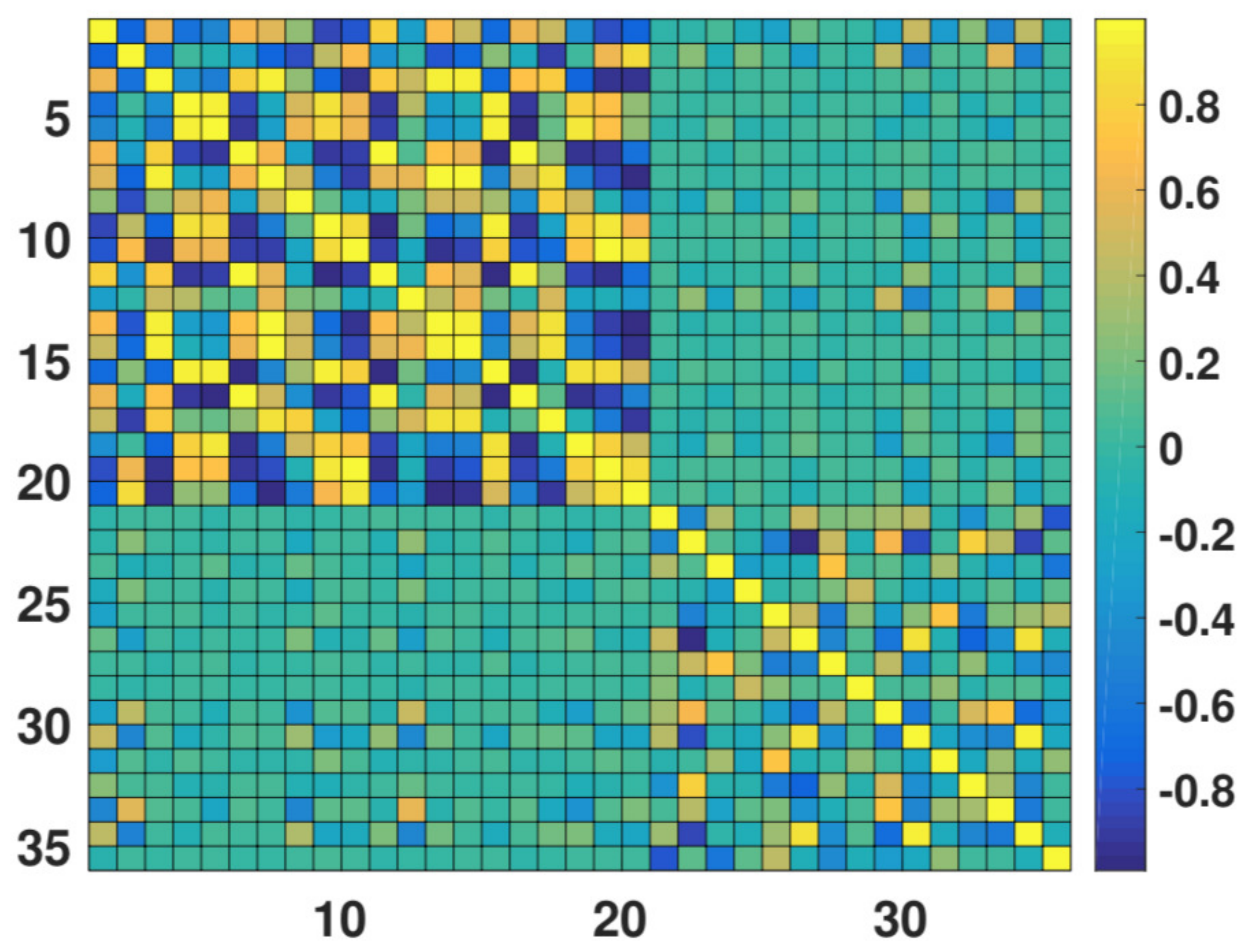} &
\includegraphics[clip=true,width=0.32\columnwidth]{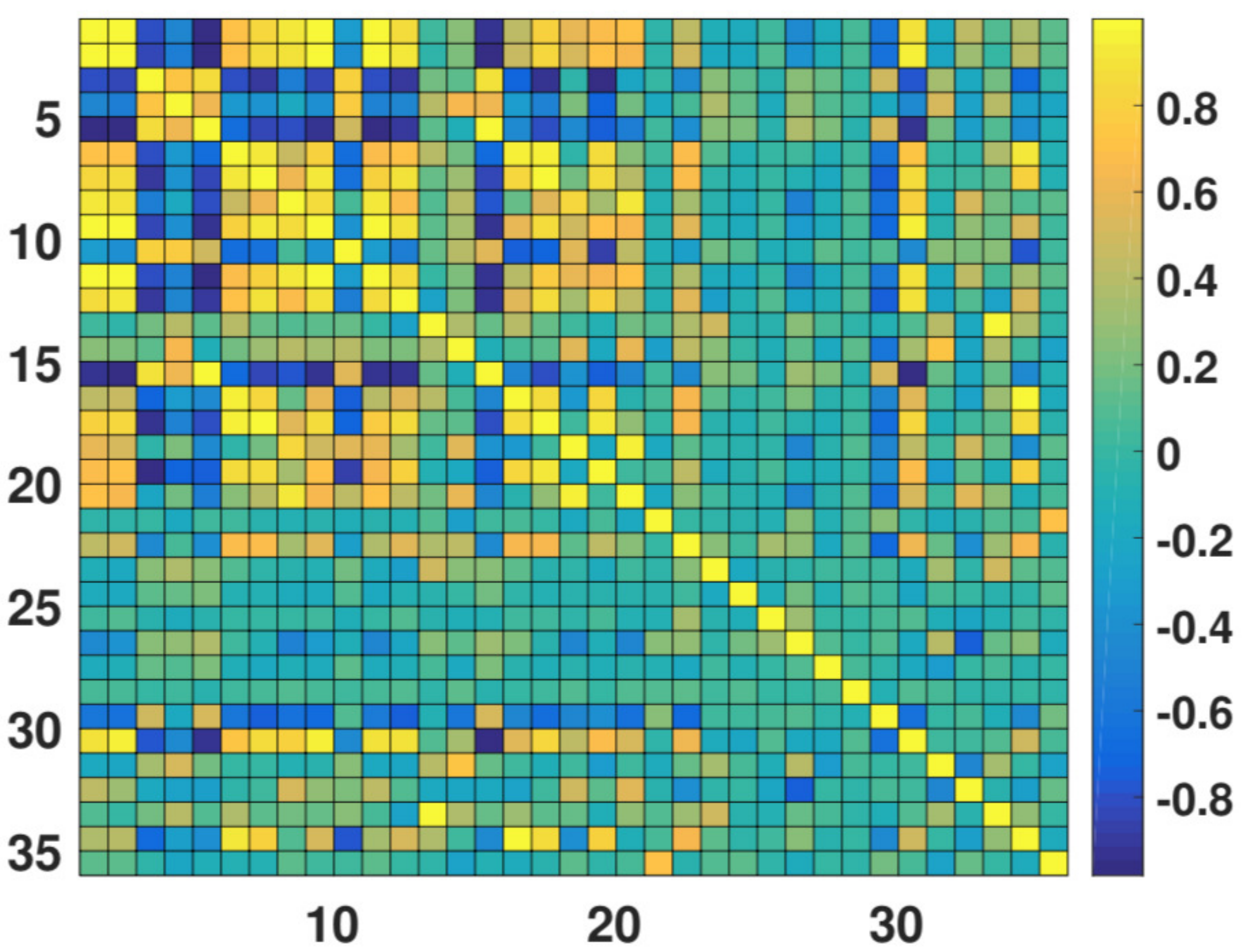}
\end{tabular}
\end{figure}

Figure~\ref{fig:6} displays the correlation matrices of the EnKF after $1$ year of assimilation. The ensemble size is $N=15$ members, observations of the full system are assimilated every $\Delta t^{\rm ocn}=\Delta t^{\rm atm}=24$hrs. MAOOAM configurations $36wk$ and $36st$ are used, and in the latter case the assimilation experiments are initialised alternatively in the active (panel (b)) and passive (panel (c)) regimes (see also Fig.~\ref{fig:2}). The figure clearly reveals the impact of the coupling on the correlation between atmosphere (top-left $20\times20$ portion) and the ocean (bottom-right $16\times16$ portion): when the coupling is weak (panel (a)) the off-diagonal entries are very small, and emerge when the coupling is increased in configuration $36st$. As expected, the atmosphere-ocean correlation is much larger in the passive regime; it is in fact the effect of the ocean that dominates in this regime and is reflected in cross-correlation patterns. It is remarkable that an ensemble of as few as $15$ members in the EnKF is able to provide physically-sound correlation patterns.

\begin{figure}
\begin{center}
\caption{EnKF analysis RMSE averaged over $300$ years for the six model configurations. The system is fully observed every $24$hrs. The number of non-negative LEs of each of the model configurations is indicated by the vertical dashed lines.}
\label{fig:7}
\includegraphics[clip=true,width=0.75\columnwidth]{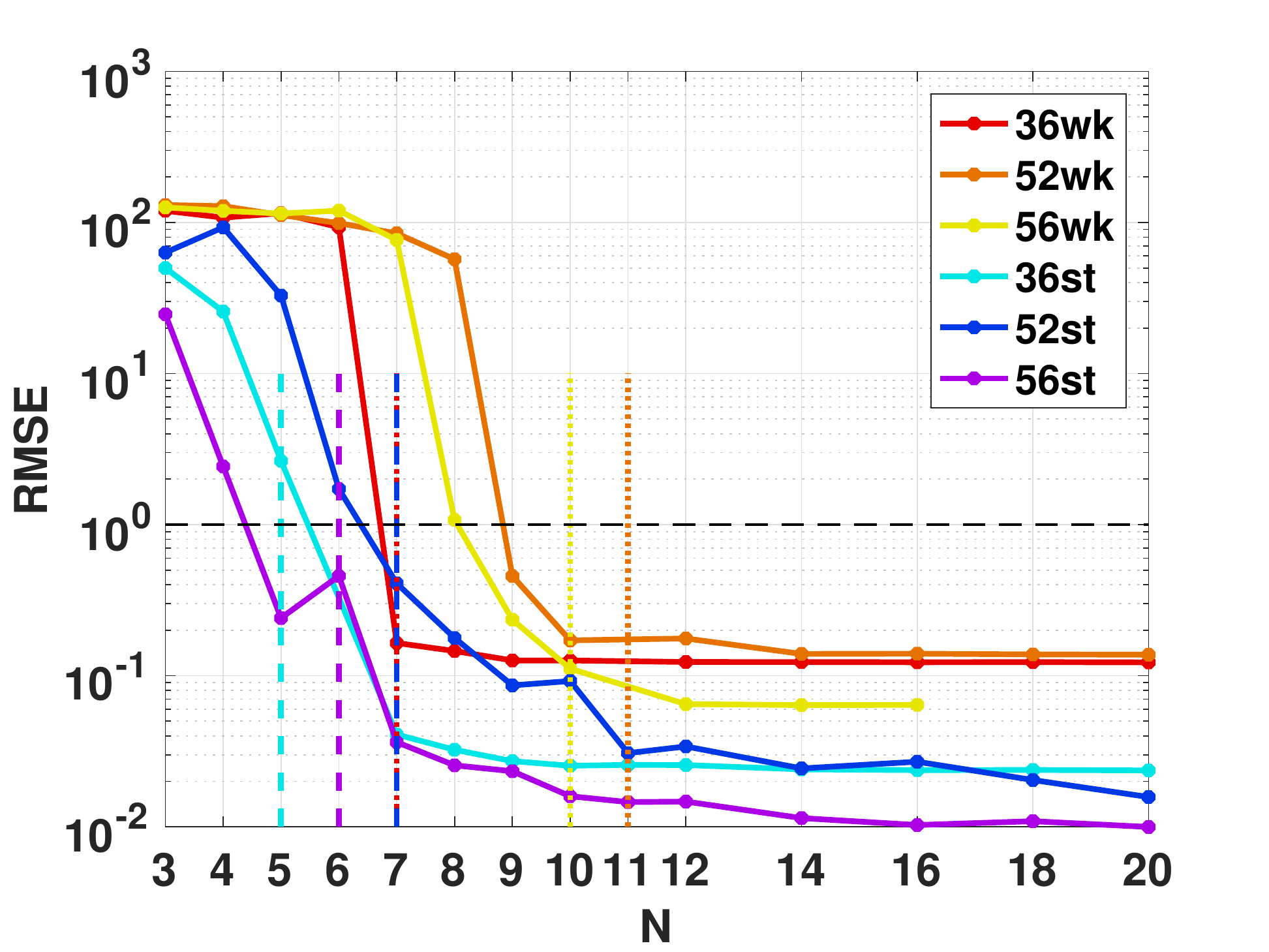}
\end{center}
\end{figure}

The relation between ensemble size and skill (in terms of the RMSE of the EnKF analysis) is studied in Fig.~\ref{fig:7}. The figure shows the global RMSE (over the whole model's domain) time averaged over $300$ years as a function of the ensemble size $N$. The RMSE is normalised with the standard deviation of the observational error, so that it has to be lower than $1$ for the EnKF to be performing satisfactorily. 
Observation type and frequency are the same as for Fig.~\ref{fig:6}: the full system is observed every $24$hrs.  

The figure shows clearly how the RMSE of the EnKF analysis decreases below the observational error level, as soon as the number of members exceeds the number of non-negative LEs. This number is indicated by the vertical dashed lines for all of the model configurations (cf. Tab.~\ref{tab:3}). Together with \cite{haussaire2016}, this result confirms, and extends to coupled dynamics, what is described in \cite{bocquet2017four} for system having a single scale of motion. This behaviour is due to the fact that, when the system is sufficiently well observed, the error dynamics behaves quasi-linearly and the errors are confined within the unstable subspace of the system. As soon as the ensemble subspace is able to fully align to the unstable subspace, the EnKF effectively reduces the error. Importantly, Fig.~\ref{fig:7} implies that, even in coupled systems, when the aforementioned condition on the observations holds, a deterministic EnKF will only need to have a number of ensemble members larger than $n_0$ to achieve good performance. Nevertheless, as opposed to the behaviour of uncoupled systems, we observe here a gradual error reduction in some cases, {\it e.g.} $52st$ and $56wk$, even for $N>n_0$. This behaviour was already observed by \cite{Penny-2019} and conjectured to be due to the extended spectrum of near‐zero LEs in the coupled system.
In fact, these quasi-neutral asymptotic LEs have high probability to be instantaneously positive. It is therefore preferable (if not mandatory) to have them accounted for in the EnKF update, so as to counteract the upwell of unfiltered error from asymptotically weakly stable (but often locally unstable) directions, as explained in \cite{grudzien2018}.

\begin{figure}
\begin{center}
\caption{Time series of the RMSE of the EnKF analysis and the ensemble spread for the configurations $36wk$ and $36st$, for the four class of model variables, atmosphere and ocean temperature and streamfunction.}
\label{fig:8}
{\bf (a) 36wk} 
\includegraphics[clip=true,width=1.1\columnwidth]{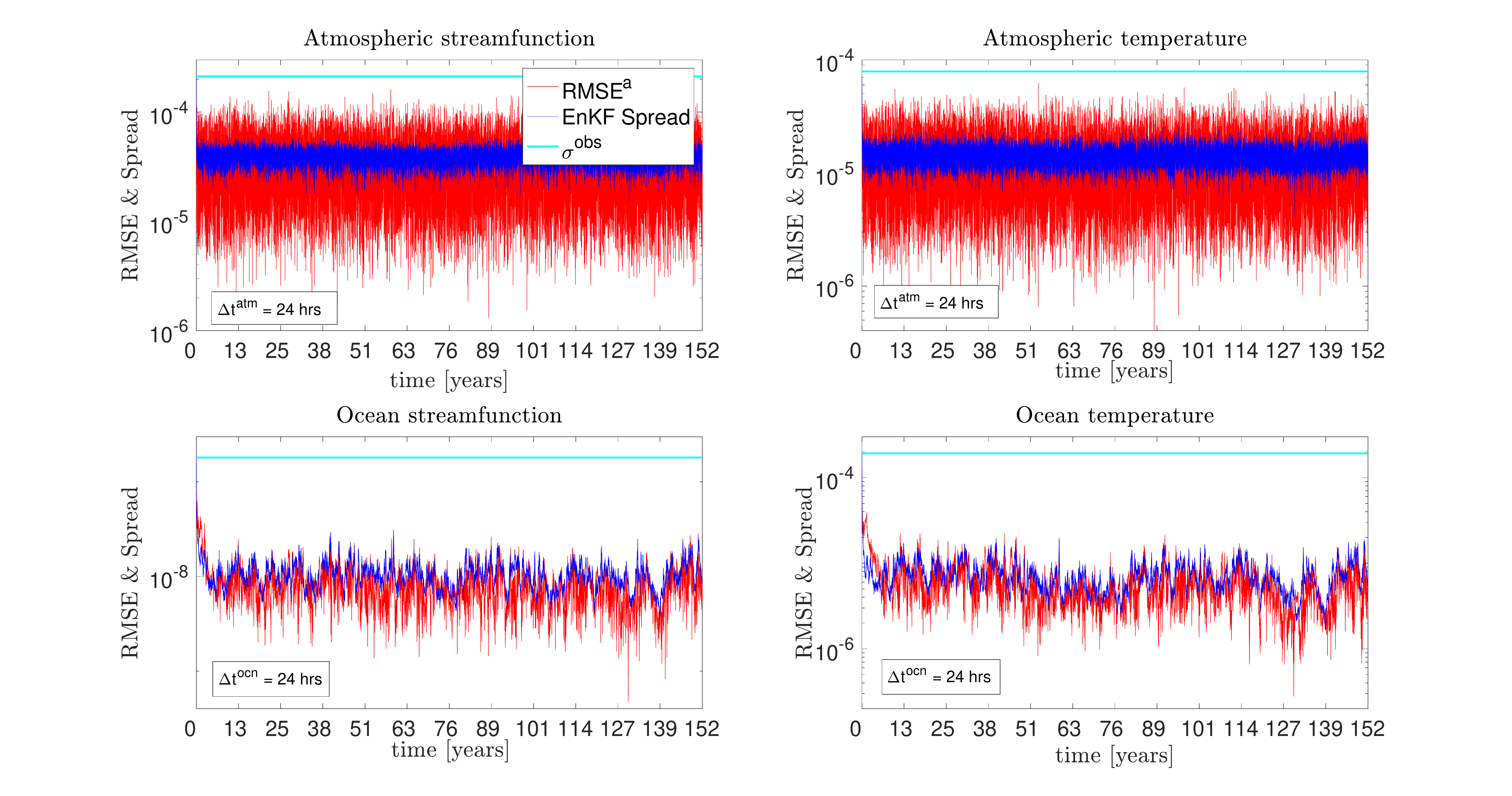} \\
{\bf (b) 36st}  
\includegraphics[clip=true,width=1.1\columnwidth]{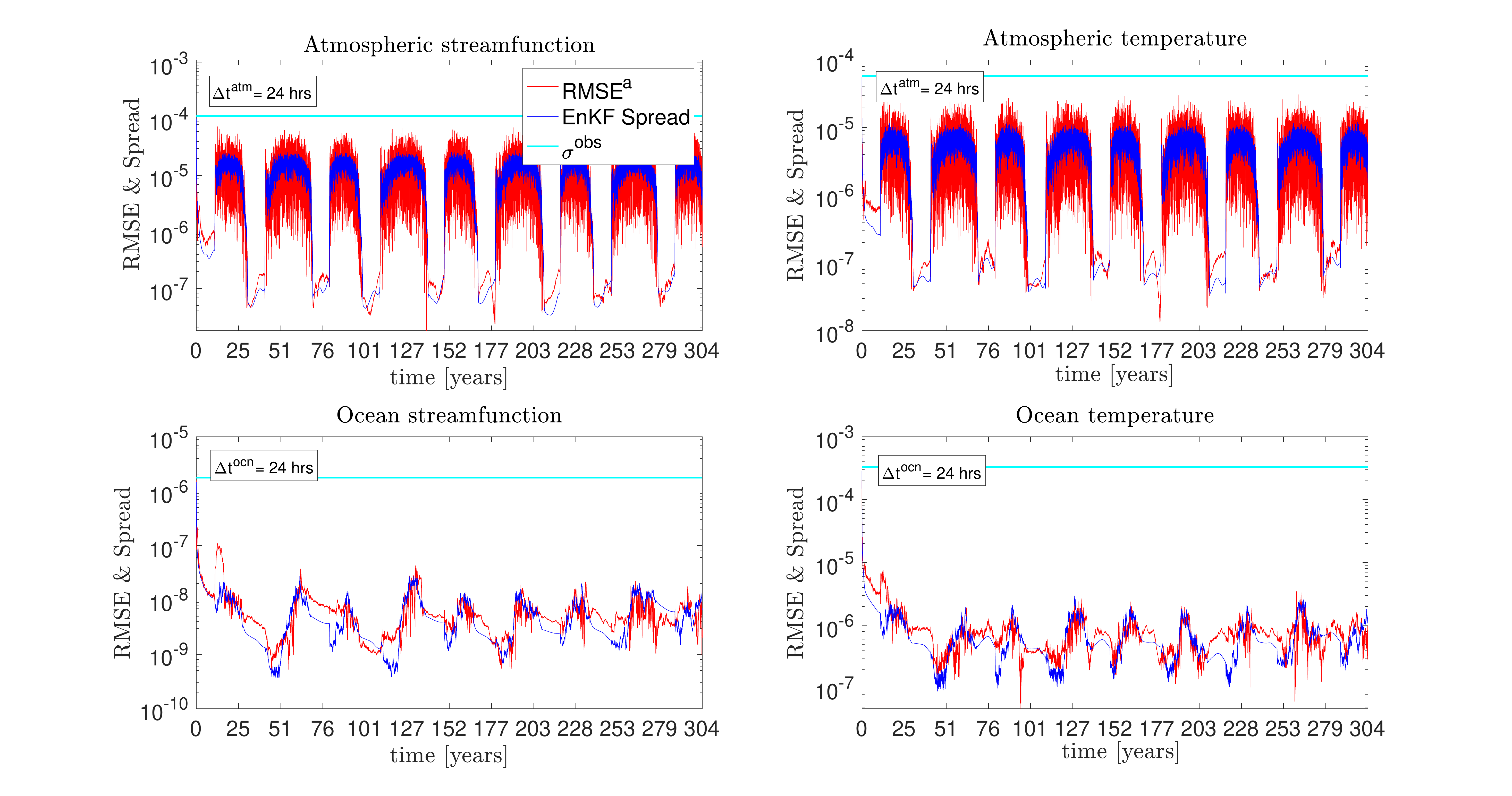} 
\end{center}
\end{figure}

Figure~\ref{fig:8} shows the time series of the RMSE of the analysis together with the ensemble spread for the configurations $36wk$ and $36st$, for the four variables; $N=15$ and $\Delta t^{{\rm ocn}}=\Delta t^{{\rm atm}}=24$hrs. Note that the experiment in the configuration $36st$ lasts for twice the duration of the $36wk$; this choice is done to balance for the slower time scale of the system when the coupling is stronger. The observational error level is also displayed for reference. 

As anticipated from Fig.~\ref{fig:7}, in all cases the RMSE is well below the observational error. The ensemble spread is also consistent with the RMSE, proving the sound functioning of the EnKF. The different temporal scales between atmosphere and ocean, as well as between weak and strong coupling configurations are evident. The figure also highlights the switch between the active and passive regimes in the $36st$ configuration. It is remarkable how well the EnKF is able to adjust to them and properly estimate the state. 

\begin{figure}
\begin{center}
\caption{RMSE of the EnKF analysis as a function of (a) the atmospheric and ocean data interval, $\Delta t^{{\rm atm}}=\Delta t^{{\rm ocn}}$; (b) atmospheric data interval, $\Delta t^{{\rm atm}}$, (c) ocean data interval, $\Delta t^{{\rm ocn}}$. In case (b) only the atmospheric data are assimilated whereas in case (c) only the ocean data are assimilated. Model configuration $36wk$ and $N=15$.}
\label{fig:9}
\begin{tabular}{l}
\\
{\bf (a) Observations: Atmosphere and Ocean} \\
\includegraphics[clip=true,trim={6.5cm, 1cm, 6cm, 1cm},width=1\columnwidth]{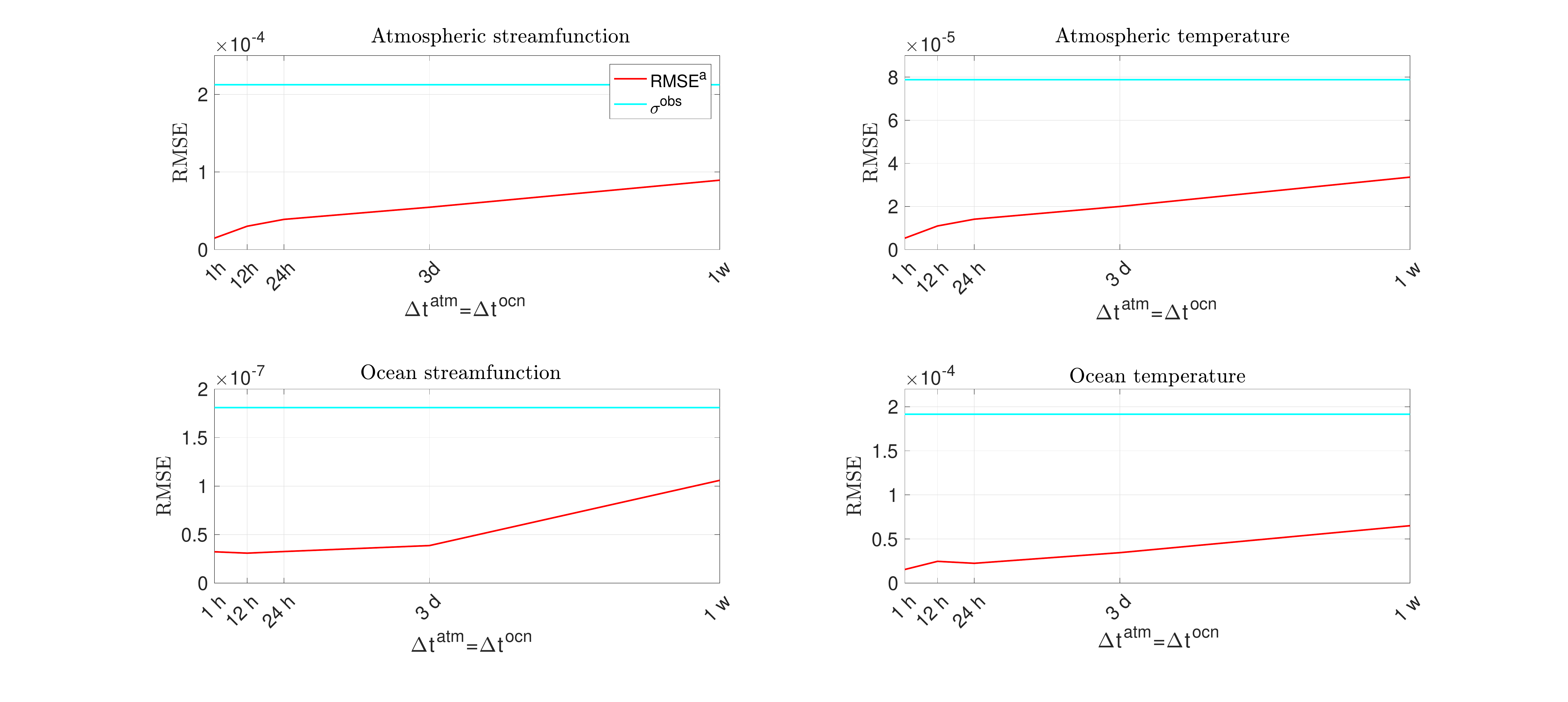} \\
{\bf (b) Observations: Atmosphere} \\
\includegraphics[clip=true,trim={6.5cm, 1cm, 6cm, 1cm},width=1\columnwidth]{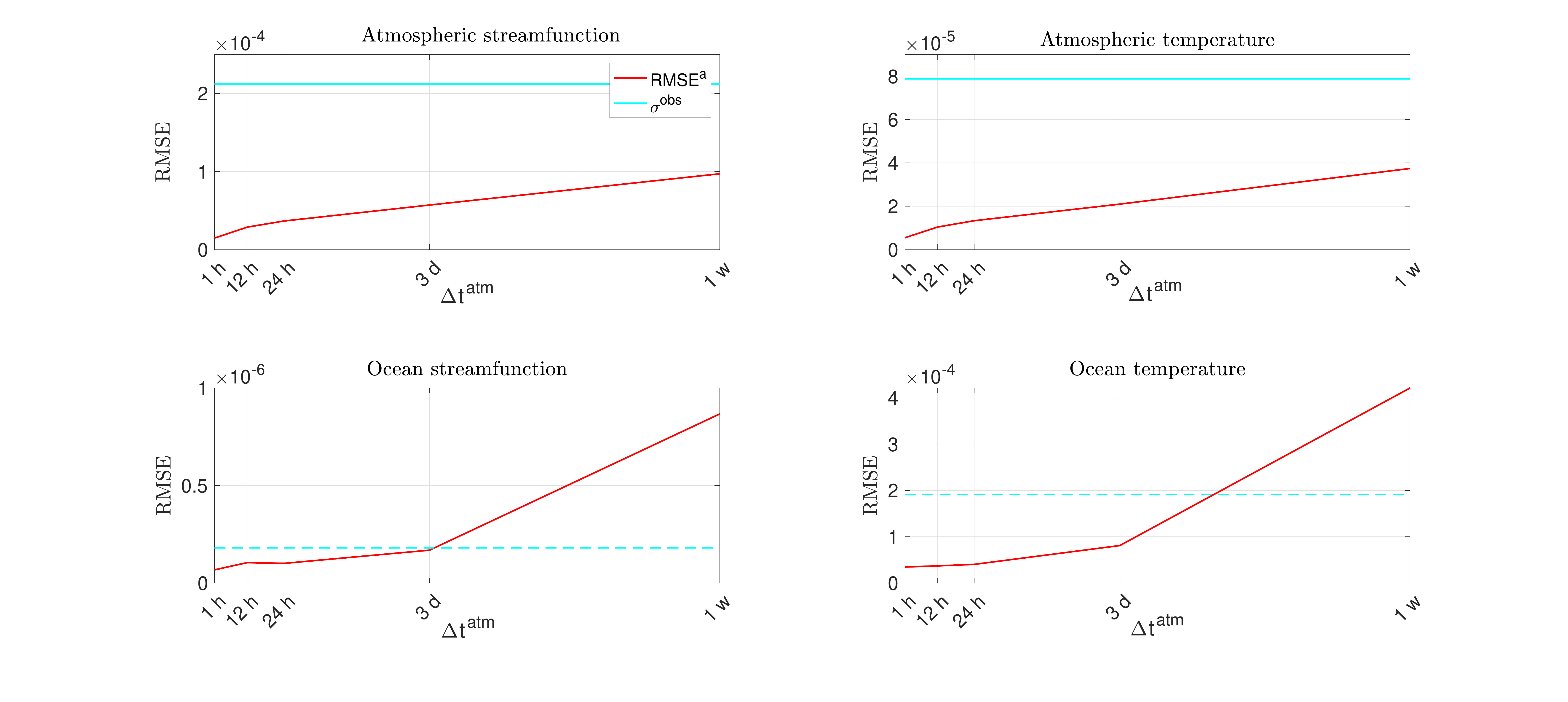}\\
{\bf (c) Observations: Ocean} \\
\includegraphics[clip=true,trim={6.5cm, 1cm, 6cm, 1cm},width=1\columnwidth]{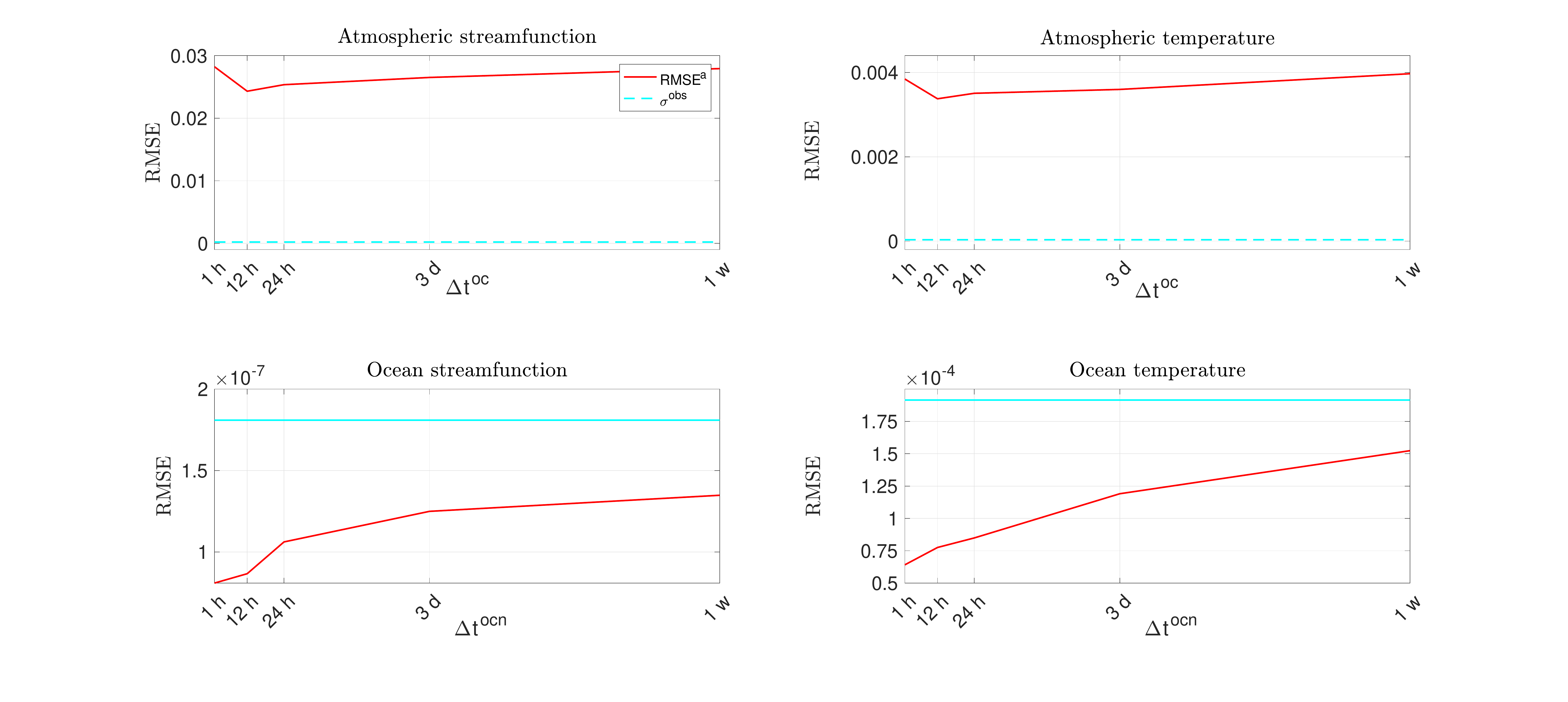}
\end{tabular}
\end{center}
\end{figure}

The effect of changing the observational intervals is studied in Fig.~\ref{fig:9} which shows the RMSE of the EnKF analysis as a function of $\Delta t^{{\rm ocn}}$ and $\Delta t^{{\rm atm}}$, for the configuration $36wk$. Results (not shown) for the other configurations and coupling are qualitatively equivalent. Experiments last $1$ year, the ensemble size is $N=15$ members and errors are averaged both in time (over this $1$ year) and on a sample of $10$ initial conditions. In Fig.~\ref{fig:9}a, atmosphere and ocean are both and simultaneously observed, but in Fig.~\ref{fig:9}b and Fig.~\ref{fig:9}c, only the atmosphere or the ocean is observed, respectively. 

When the atmosphere and the ocean are both observed (panels (a)), the RMSE shows a monotonic growth trend when $\Delta t^{{\rm ocn}}$ and $\Delta t^{{\rm atm}}$ are increased, although the RMSE of the ocean streamfunction, and to a lesser extent the ocean temperature, seem quite insensitive in the interval $1h\le\Delta t\le 3d$. 
Note that the RMSEs of all four variables stay below the observational error level for all the considered observational intervals.    
The situation changes slightly when only atmospheric data are used (panels (b)). Remarkably the error level in the atmosphere appears insensitive to the removal of the ocean observations. While this is obviously not the case for the ocean RMSE, which in fact increases when only atmospheric data are available, it is interesting to observe that it only increases very little. In practice, when $1h\le\Delta t^{{\rm atm}}\le3d$, the ocean RMSE stays below the observational level even in the absence of ocean data; all information is brought and propagated from the atmosphere. 
The importance of atmospheric data is further highlighted by the results in panels (c) in which atmospheric observations are removed and the EnKF only assimilates ocean data. We see how the atmospheric RMSE is now above the observational level consistently in all variables. Ocean RMSE, while slightly lower than when only atmospheric data were available (panels (b)), it is not as low as when both ocean and atmosphere were observed (panels (a)).
The importance of observing the fast scale, as conjectured in Sect.~\ref{sec:intro3}, is thus corroborated by these numerical findings.
It is also relevant to observe that, despite its slow time scale, the ocean analysis improves even when observations are assimilated as frequently as $1h$ and $12h$, in line with what found in \cite{Penny-2019}. 

\begin{figure}
\caption{RMSE of the EnKF analysis as a function of (a) the atmospheric data interval with fixed ocean data interval, $\Delta t^{{\rm ocn}}=1$w; (b) the ocean data interval with fixed atmospheric data interval, $\Delta t^{{\rm atm}}=12$h. The model configuration is $36wk$, with an ensemble size set to $N=15$.} 
\label{fig:10}
\begin{tabular}{l}
\\
{\bf (a) Varying/Fixed Atm/Oc data freq} \\
\includegraphics[clip=true,trim={6.5cm, 1cm, 6cm, 1cm},width=1\columnwidth]{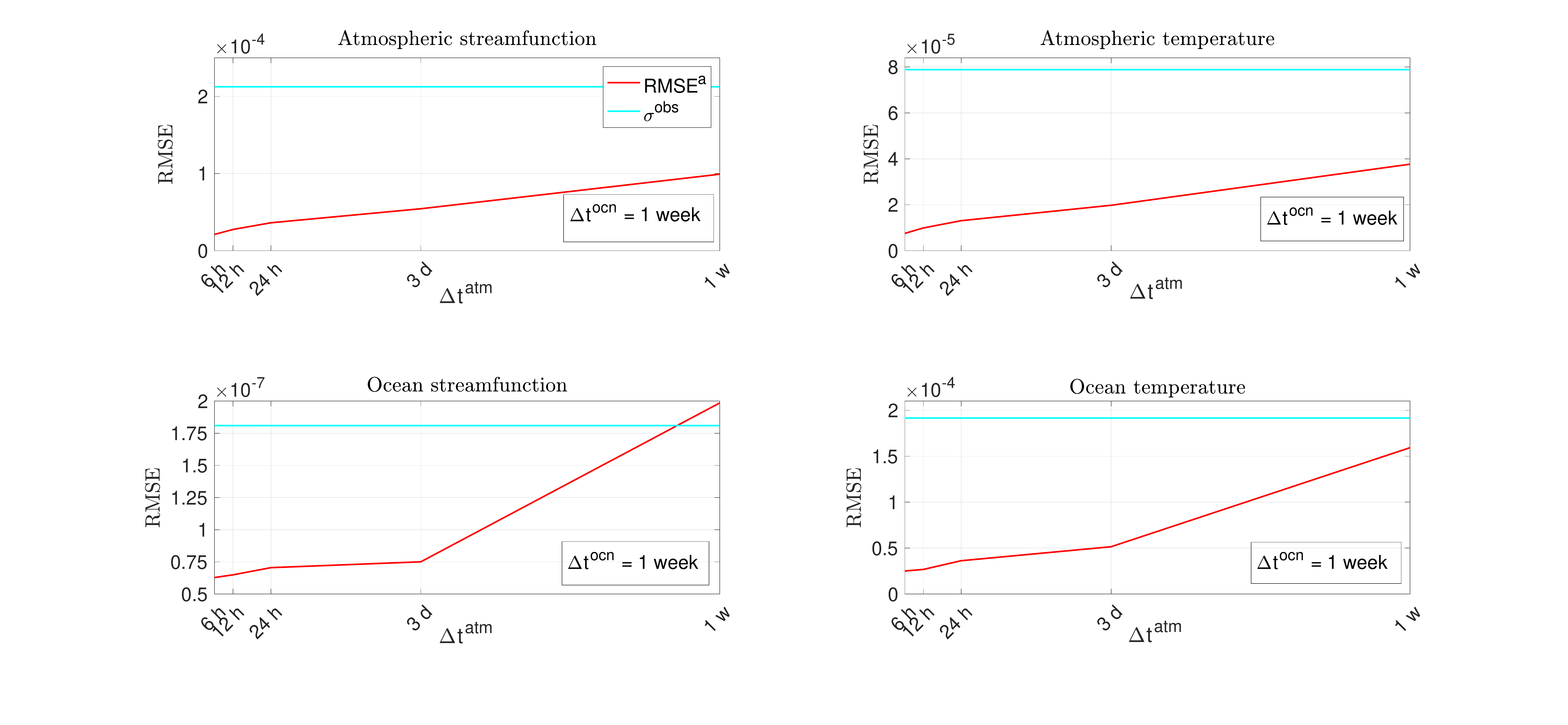} \\
{\bf (b) Varying/Fixed Oc/Atm data freq} \\
\includegraphics[clip=true,trim={6.5cm, 1cm, 6cm, 1cm},width=1\columnwidth]{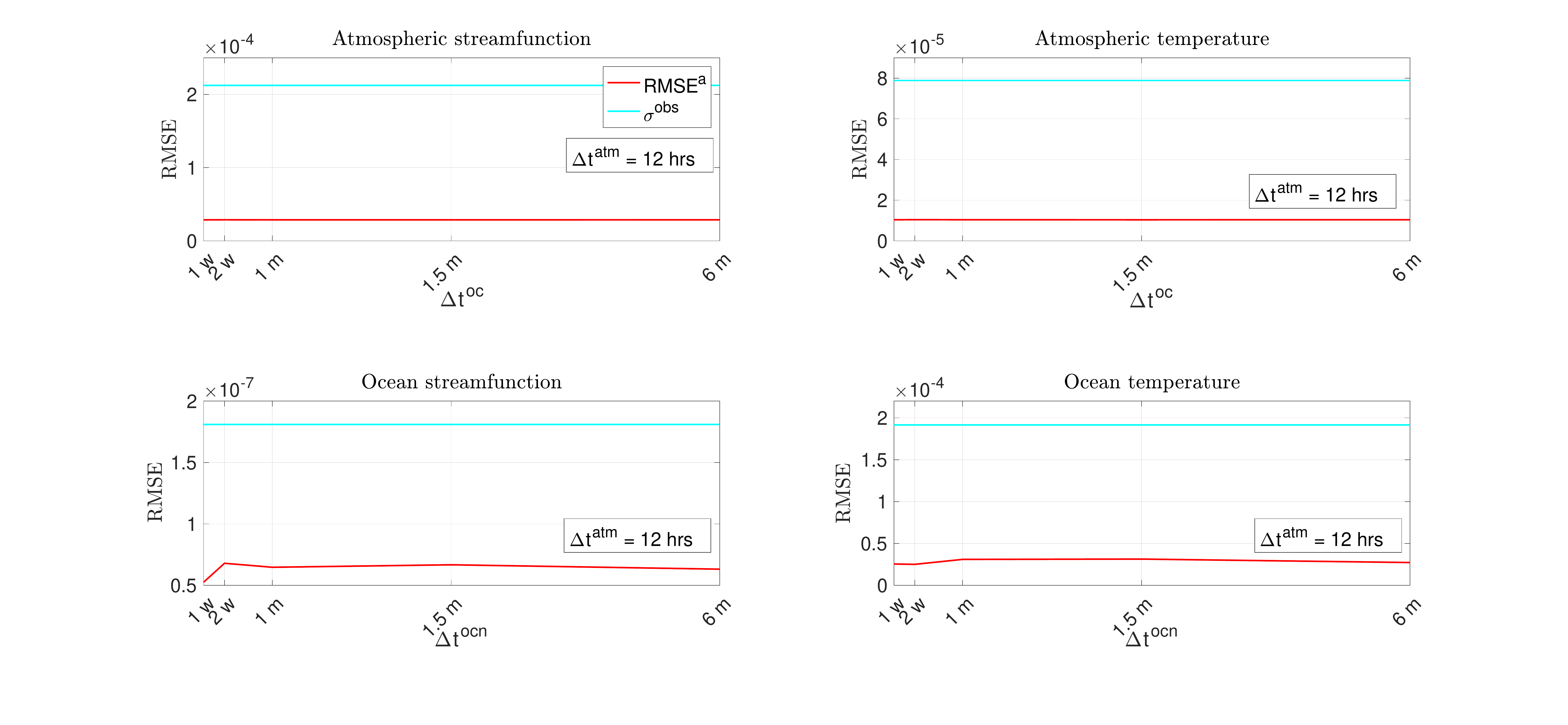}
\end{tabular}
\end{figure}

Figure~\ref{fig:10} is similar to Fig.~\ref{fig:9} except that now the observation interval is kept fixed in one component, either atmosphere or ocean, while varied in the other. In the experiments of panels (a) the ocean data are assimilated every week, $\Delta t^{{\rm ocn}}=1$w, while the frequency of atmospheric data is changed. Conversely, panels (b) show experiments where the atmospheric data are assimilated every half-day, $\Delta t^{{\rm atm}}=12$hrs, while the frequency of ocean data is changed. Similarly to Fig.~\ref{fig:9}, results of Fig.~\ref{fig:10} also confirms the importance of providing a sufficient control of the fast scale (the atmosphere) error growth, by keeping the observation frequency high enough. In fact, the comparison of panels (a) and (b) reveal how in the latter case, when ocean data frequency is changed but the atmosphere is observed every $\Delta t^{{\rm atm}}=12$hrs the analysis RMSE is maintained consistently below the observational error. This contrasts with the behaviour shown in panels (a) where, although the ocean is observed every $\Delta t^{{\rm ocn}}=1$w, the analysis RMSE in all variables grows with the increase of the atmospheric data frequency, eventually reaching a level higher than the observational error.  

\section{Conclusion}
\label{sec:concl}

The term ``coupled data assimilation'' (CDA) has been increasingly used in recent years to refer to the application of data assimilation (DA; {\it e.g.} \cite{carrassi2018data}) in dynamical systems possessing many, and separated, scales of motion that are coupled together in their dynamical equations \cite{penny2017coupled}. Systems of this sort are common in many areas of sciences, but CDA has emerged distinctively in climate science where Earth system numerical models couple together models of the atmosphere, land, ocean and cryosphere. Classical DA is prone not to work efficiently because the scale separation acts as a barrier hindering the transmission of the information content across model components ({\it e.g.} ocean and atmosphere). 
Understanding origins and causes limiting classical DA is important and may help guiding adaptations and novel solutions.

We have provided an introduction to CDA in Sect.~\ref{sec:intro}, together with a survey of the current status of the research in the field. By using dynamical arguments, in Sect.~\ref{sec:intro3} we traced back the core issue and illustrated in which way, to a first order, information flows from the fast to the slow scale or vice-versa. Furthermore we conjectured how observations of both scales have to be temporally distributed in order to best reduce the state estimation error. We deduced that: (i) cross components effects are generally stronger in the direction from the slow to the fast scale, so that observations of the slow scale may benefit to the fast, but, (ii) intra-component effects are much stronger in the fast scale. The fast scale must be controlled by frequently enough observations to prevent the error to grow up and affect the slow scale. 

The above is in overall agreement with previous works that, while having shown benefit in both directions, have also indicated atmospheric data to be more effective in constraining the ocean than the opposite (see {\it e.g.} \cite{Penny-2019} and references therein). This includes studies (see {\it e.g.} \cite{smith2015exploring} and references therein) where uncoupled but forced models of the atmosphere or the ocean are considered. 
In cases when the ocean is forced with pre-computed atmospheric surface fluxes, error in the latter are responsible for biases in the ocean, revealing an atmosphere-to-ocean impact; see also \cite{Penny-2019} for an extensive discussion on the transition from uncoupled-forced models to weakly- and strongly-coupled DA.

Our conjectures have been confirmed in numerical experiments performed with the modular arbitrary-order coupled atmosphere-ocean model, MAOOAM \cite{DeCruz_et_al_2016}, in which a state-of-the-art ensemble Kalman filter, the EnKF-N \cite{bocquet2015} has been implemented. MAOOAM has been used in six different configurations, having different sizes and atmosphere-ocean coupling strengths. The attractors in the weak and strong coupling cases appear very different, with the strong one showing two distinct regimes and a low frequency variability with a period of about $16$ years.   
We have characterised the model stability properties via the spectrum of Lyapunov exponents, the Kolmogorov entropy, the Kaplan-Yorke attractor dimension and the covariant Lyapunov vectors (CLVs). In particular, the averaged projections of the CLVs onto the state vector reveal how the different model instabilities are driven by the atmosphere and/or the ocean.   

The experiments with the EnKF-N have confirmed the behaviour anticipated in Sect.~\ref{sec:intro3}: atmosphere has to be observed frequently enough ({\it i.e.} about every $12$hrs) in order to achieve a global analysis (including the ocean) with low error ({\it i.e.} below the observational error). Moreover, experiments largely prove that, likewise uncoupled dynamics \cite{bocquet2017four}, deterministic EnKFs (to which category the EnKF-N belongs) require an ensemble at least as large as the number of non-negative Lyapunov exponents (assuming localisation is not used), to get satisfactory results in coupled systems too. 
However, as opposed to uncoupled systems, 
whenever the model displays a degeneracy-like in the Lyapunov spectrum (with many near-zero exponents)
the analysis error still gradually decreases for $N>n_0$; in uncoupled systems the analysis error reduction almost fully ceases when $N>n_0$. 
This behaviour, originally observed in \cite{Penny-2019}, is arguably due to the presence of multiple near-neutral asymptotic directions, with a high chance to be locally unstable. 

Although asymptotically neutral or weakly-stable, these directions may display high variance in the local error growth rate, thus be often intermittently unstable. As rigorously proved by \cite{grudzien2018JUQ,grudzien2018} this situation is known to drive the error upwell from the unfiltered to the filtered subspace, eventually leading to divergence. 
It seems thus paramount that the EnKF ensemble subspace encompasses at least all of these near-neutral directions, preferably also the asymptotically weakly stable.. 
Furthermore, given that these directions are generated by the coupling itself (see Fig.~\ref{fig:6} and \cite{vannitsem2016statistical}) their impact on the performance of the EnKF in coupled systems is expected to be ubiquitous. Along these lines, using an idealised multi-scale system made up of coupled copies of the Lorenz 3-variables model, the authors of \cite{npg-27-51-2020} demonstrate that weakly stable directions are needed in situations of strong nonlinear dynamics and intermittent error growth. 
Similarly, the authors of \cite{lucarini2020new} have suggested that the variability in the number of unstable modes associated to unstable periodic orbits in a simple Earth system model, can explain the observed very different predictability of individual atmospheric blocking events, and have argued that DA must thus cautiously incorporate stable modes too.   

One of our current research endeavours is the study of suitable reduced-rank formulations of the EnKF that take into account the unstable modes in coupled systems, in analogy to the assimilation in the unstable subspace (AUS; \cite{palatella2013a}) so far applied to uncoupled systems. Similarly, the map between the instability rank and the state vector drawn by the CLVs may help designing monitoring strategies in which observing devices are deployed in the areas of large CLVs (see {\it e.g.} \cite{carrassi2007} for a similar strategy based on breeding vectors of the DA cycle). Part of the questions related to extending AUS to coupled dynamics have been undertaken in the recent work \cite{npg-27-51-2020}, although many still remain to be addressed using more realistic models with the aforementioned Lyapunov degeneracy.    
This can be done using MAOOAM given that its number of positive and neutral exponents can be very large \cite{npg-25-387-2018}, as they manifest the coupling mechanisms. Do we still need such a large amount of ensemble members, or is there a limit beyond which a further increase is not necessary anymore and, if so, under which circumstances? These questions are worth addressing to properly set up the EnKF in multi-scale dynamics.

\begin{acknowledgements}
The numerical experiments, and their analysis, were performed during the internship that Maxime Tondeur spent at NERSC and RMI in 2017, as part of his Master at \'Ecole des Mines Paris.  
The internship was funded by the Nordic Centre of Excellence EmblA of the Nordic Countries Research Council, NordForsk. Alberto Carrassi has been funded by the Trond Mohn Foundation under the project number BFS2018TMT0 and by the Natural Environment Research Council (Agreement PR140015 between NERC and the National Centre for Earth Observation).
St\`ephane Vannitsem acknowledges partial support by the Belgian Science Policy Office under contract BR/165/A2/Mass2Ant. CEREA is a member of the Institut Pierre-Simon Laplace (IPSL).

\end{acknowledgements}

\section*{Appendix: Lyapunov exponents and covariant Lyapunov vectors}

The initial state of a system is never known exactly since the process of measurement and data assimilation is always subjected to finite precision.
To clarify the implications of the presence of such an error we consider an initial state displaced slightly from $\x(t_0)=\x_0$ by an 
initial error $\delta\x_0$.  This perturbed initial state generates a new trajectory in phase space
and we define the instantaneous error vector as the vector joining the points of the reference trajectory and the perturbed one at
a given time, $\delta\x_t$. Provided that this perturbation is sufficiently small, its dynamics
can be described by the linearised equation,
\begin{equation}
\frac{{\rm d}\delta\x}{{\rm d}t} =  \frac{{\rm \partial }\f}{{\rm \partial}\x}_{\vert \x_t}\delta\x
\label {linear}
\end{equation}
and the formal solution can be written as,
\begin{equation}
\delta\x_t = {\bf M}_{t:t_0}(\x_0) \delta \x_0
\end{equation}
where the matrix $\bf{M}$, referred as the resolvent matrix, plays an important role in error growth dynamics
as revealed when writing the Euclidean norm of the error,
\begin{eqnarray}
E_t & = & \|\delta \x_t \|^2 = \delta\x_t^{\rm T} \delta\x_t  \nonumber \\ 
 & = & \delta\x_0^{\rm T}\bM_{t:t_0}^{\rm T}(\x_0)\bM_{t:t_0}(\x_0)\delta\x_0
\label{eucl}
\end{eqnarray}
The growth of $E_t$ is conditioned by the eigenvalues of the matrix
$\bM^{\rm T}\bM$, where $(.)^{\rm T}$ indicates transposition (and complex conjugation if necessary).
In ergodic theory of chaotic systems, the double
limit of infinitely small initial errors and infinitely long times, is usually considered ({\it e.g} \cite{Eckmann1985}). In
these limits the divergence of initially closed states is determined by the logarithm of the eigenvalues of the
matrix $[\bM^{\rm T}\bM]^{2(t-t_0)}$ that are referred to as the Lyapunov exponents (LEs).
The full set of LEs of a system is called the Lyapunov spectrum which are usually represented in
decreasing order. 

Associated with each of these exponents a natural direction of (in)stability can be defined which is a local property on the
attractor of the system, these are known as the covariant Lyapunov vectors (CLVs).
These CLVs were first introduced in \cite{Ruelle1979} and later discussed in \cite{Legras1995,Trevisan1998}.  
They form a norm-independent and covariant basis of the tangent 
linear space, providing a splitting between the unstable manifold. This splitting describes the unstable perturbations leading to the divergence of the 
trajectories, the neutral manifold, typically associated with the direction of the flow, and the stable manifold, which corresponds to the 
contracting directions. Several algorithms for computing these vectors are available \cite{Ginelli2007,Wolfe2007,Kuptsov2012,Froyland2013}. 

The CLVs are defined through a suitable geometric construction involving both the orthogonal forward and backward Lyapunov vectors, whose computation can be 
seen as a byproduct of the usual Benettin et al. algorithm \cite{Benettin80} used for estimating of LEs, see also the nice reviews on these topics in 
\cite{Eckmann1985,Kuptsov2012,Froyland2013}. The size of the 
perturbations oriented according to the CLVs grows or decays with an approximate exponential law, where the average of the fluctuating rates of 
growth or decay correspond one-to-one to the LEs. As opposed to the CLVs, the forward and backward Lyapunov vectors (except for the first) are not covariant, so that it 
is hard to interpret them physically \cite{Pazo2010}. Therefore, CLVs allow for associating a time-dependent field to each LE, thus providing a 
connection between observed rates of growth and decay of perturbations and the corresponding physical modes of the system. 

A recent analysis of the statistical and dynamical properties of these vectors in the context of MAOOAM has been performed 
in \cite{vannitsem2016statistical}. A remarkable result is the splitting of the tangent space in two categories of CLVs, a first set mostly associated
with the dynamics within the atmosphere, the most unstable and stable directions, and a second set of vectors for which  the corresponding Lyapunov
exponents are close to $0$. The latter set forms a quasi-neutral subspace of truly coupled ocean-atmosphere modes.


\begin{thebibliography}{10}
\providecommand{\url}[1]{{#1}}
\providecommand{\urlprefix}{URL }
\expandafter\ifx\csname urlstyle\endcsname\relax
  \providecommand{\doi}[1]{DOI~\discretionary{}{}{}#1}\else
  \providecommand{\doi}{DOI~\discretionary{}{}{}\begingroup
  \urlstyle{rm}\Url}\fi

\bibitem{asch2016data}
Asch, M., Bocquet, M., Nodet, M.: Data {A}ssimilation: {M}ethods, {A}lgorithms,
  and {A}pplications.
\newblock Fundamentals of Algorithms. SIAM, Philadelphia (2016)

\bibitem{bach2019local}
Bach, E., Motesharrei, S., Kalnay, E., Ruiz-Barradas, A.: Local
  atmosphere--ocean predictability: Dynamical origins, lead times, and
  seasonality.
\newblock Journal of Climate \textbf{32}(21), 7507--7519 (2019)

\bibitem{ballabrera2009data}
Ballabrera-Poy, J., Kalnay, E., Yang, S.C.: Data assimilation in a system with
  two scales—combining two initialization techniques.
\newblock Tellus A \textbf{61}(4), 539--549 (2009)

\bibitem{Benettin80}
Benettin, G., Galgani, L., Giorgilli, A., Strelcyn, J.M.: Lyapunov
  characteristic exponents for smooth dynamical systems and for hamiltonian
  systems; a method for computing all of them. part 1: Theory.
\newblock Meccanica \textbf{15}(1), 9--20 (1980)

\bibitem{bocquet2011}
Bocquet, M.: Ensemble {K}alman filtering without the intrinsic need for
  inflation.
\newblock Nonlinear Proc. Geoph. \textbf{18}, 735--750 (2011).
\newblock \doi{10.5194/npg-18-735-2011}

\bibitem{bocquet2017four}
Bocquet, M., Carrassi, A.: Four-dimensional ensemble variational data
  assimilation and the unstable subspace.
\newblock Tellus A \textbf{69}(1), 1304504 (2017)

\bibitem{bocquet2015}
Bocquet, M., Raanes, P.N., Hannart, A.: Expanding the validity of the ensemble
  {K}alman filter without the intrinsic need for inflation.
\newblock Nonlinear Proc. Geoph. \textbf{22}, 645--662 (2015).
\newblock \doi{10.5194/npg-22-645-2015}

\bibitem{browne2019weakly}
Browne, P.A., de~Rosnay, P., Zuo, H., Bennett, A., Dawson, A.: Weakly coupled
  ocean--atmosphere data assimilation in the ecmwf nwp system.
\newblock Remote Sensing \textbf{11}(3), 234 (2019)

\bibitem{brunet2015seamless}
Brunet, G., Jones, S., Ruti, P.M., et~al.: Seamless prediction of the Earth
  System: from minutes to months.
\newblock World Meteorological Organization (2015)

\bibitem{brunet2010collaboration}
Brunet, G., Shapiro, M., Hoskins, B., Moncrieff, M., Dole, R., Kiladis, G.N.,
  Kirtman, B., Lorenc, A., Mills, B., Morss, R., et~al.: Collaboration of the
  weather and climate communities to advance subseasonal-to-seasonal
  prediction.
\newblock Bulletin of the American Meteorological Society \textbf{91}(10),
  1397--1406 (2010)

\bibitem{carrassi2018data}
Carrassi, A., Bocquet, M., Bertino, L., Evensen, G.: Data assimilation in the
  geosciences: An overview of methods, issues, and perspectives.
\newblock Wiley Interdisciplinary Reviews: Climate Change \textbf{9}(5), e535
  (2018)

\bibitem{carrassi2007}
Carrassi, A., Trevisan, A., Uboldi, F.: Adaptive observations and assimilation
  in the unstable subspace by breeding on the data-assimilation system.
\newblock Tellus A \textbf{59}, 101--113 (2007)

\bibitem{carrassi2009}
Carrassi, A., Vannitsem, S., Zupanski, D., Zupanski, M.: The maximum likelihood
  ensemble filter performances in chaotic systems.
\newblock Tellus A \textbf{61}, 587--600 (2009)

\bibitem{counillon2014seasonal}
Counillon, F., Bethke, I., Keenlyside, N., Bentsen, M., Bertino, L., Zheng, F.:
  Seasonal-to-decadal predictions with the ensemble {Kalman} filter and the
  {Norwegian Earth System Model}: {A} twin experiment.
\newblock Tellus A \textbf{66}(1), 21074 (2014)

\bibitem{DeCruz_et_al_2016}
De~Cruz, L., Demaeyer, J., Vannitsem, S.: The modular arbitrary-order
  ocean-atmosphere model: \textsc{maooam}~v1.0.
\newblock Geoscientific Model Development \textbf{9}(8), 2793--2808 (2016).
\newblock \doi{10.5194/gmd-9-2793-2016}.
\newblock \urlprefix\url{https://www.geosci-model-dev.net/9/2793/2016/}

\bibitem{npg-25-387-2018}
De~Cruz, L., Schubert, S., Demaeyer, J., Lucarini, V., Vannitsem, S.: Exploring
  the lyapunov instability properties of high-dimensional atmospheric and
  climate models.
\newblock Nonlinear Processes in Geophysics \textbf{25}(2), 387--412 (2018).
\newblock \doi{10.5194/npg-25-387-2018}.
\newblock \urlprefix\url{https://www.nonlin-processes-geophys.net/25/387/2018/}

\bibitem{dirren2005toward}
Dirren, S., Hakim, G.J.: Toward the assimilation of time-averaged observations.
\newblock Geophysical research letters \textbf{32}(4) (2005)

\bibitem{doblas2013seasonal}
Doblas-Reyes, F.J., Garc{\'\i}a-Serrano, J., Lienert, F., Biescas, A.P.,
  Rodrigues, L.R.: {Seasonal climate predictability and forecasting: Status and
  prospects}.
\newblock Wiley Interdisciplinary Reviews: Climate Change \textbf{4}(4),
  245--268 (2013)

\bibitem{Eckmann1985}
Eckmann, J.P., Ruelle, D.: Ergodic theory of chaos and strange attractors.
\newblock In: The theory of chaotic attractors, pp. 273--312. Springer (1985)

\bibitem{evensen2009}
Evensen, G.: {D}ata {A}ssimilation: {T}he {E}nsemble {K}alman {F}ilter, second
  edn.
\newblock Springer-Verlag/Berlin/Heildelberg (2009)

\bibitem{Froyland2013}
Froyland, G., H{\"u}ls, T., Morriss, G.P., Watson, T.M.: Computing covariant
  lyapunov vectors, oseledets vectors, and dichotomy projectors: A comparative
  numerical study.
\newblock Physica D: Nonlinear Phenomena \textbf{247}(1), 18--39 (2013)

\bibitem{Ginelli2007}
Ginelli, F., Poggi, P., Turchi, A., Chat{\'e}, H., Livi, R., Politi, A.:
  Characterizing dynamics with covariant lyapunov vectors.
\newblock Physical review letters \textbf{99}(13), 130601 (2007)

\bibitem{grudzien2018JUQ}
Grudzien, C., Carrassi, A., Bocquet, M.: Asymptotic forecast uncertainty and
  the unstable subspace in the presence of additive model error.
\newblock SIAM/ASA J. Uncertainty Quantification \textbf{6}(4), 1335--1363
  (2018)

\bibitem{grudzien2018}
Grudzien, C., Carrassi, A., Bocquet, M.: Chaotic dynamics and the role of
  covariance inflation for reduced rank kalman filters with model error.
\newblock Nonlinear Proc. Geoph. Disc. \textbf{2018}, 1--25 (2018).
\newblock \doi{10.5194/npg-2018-4}.
\newblock
  \urlprefix\url{https://www.nonlin-processes-geophys-discuss.net/npg-2018-4/}

\bibitem{Hannart-et-al-2016}
Hannart, A., Carrassi, A., Bocquet, M., Ghil, M., Naveau, P., Pulido, M., Ruiz,
  J., Tandeo, P.: {DADA: data assimilation for the detection and attribution of
  weather and climate-related events}.
\newblock Climatic Change \textbf{136}(2), 155--174 (2016)

\bibitem{harlim2010filtering}
Harlim, J., Majda, A.J.: Filtering turbulent sparsely observed geophysical
  flows.
\newblock Mon. Weather Rev. \textbf{138}(4), 1050--1083 (2010)

\bibitem{haussaire2016}
Haussaire, J.M., Bocquet, M.: A low-order coupled chemistry meteorology model
  for testing online and offline data assimilation schemes: {L95-GRS} (v1.0).
\newblock Geosci. Model Dev. \textbf{9}, 393--412 (2016).
\newblock \doi{10.5194/gmd-9-393-2016}

\bibitem{hunt2007}
Hunt, B., Kostelich, E.J., Szunyogh, I.: Efficient data assimilation for
  spatiotemporal chaos: {A} local ensemble transform {K}alman filter.
\newblock Physica D \textbf{230}, 112--126 (2007)

\bibitem{huntley2010assimilation}
Huntley, H.S., Hakim, G.J.: Assimilation of time-averaged observations in a
  quasi-geostrophic atmospheric jet model.
\newblock Climate dynamics \textbf{35}(6), 995--1009 (2010)

\bibitem{hutt2019data}
Hutt, A., Stannat, W., Potthast, R.: Data Assimilation and Control: Theory and
  Applications in Life Sciences.
\newblock Frontiers Media SA (2019)

\bibitem{janjic2018representation}
Janji{\'c}, T., Bormann, N., Bocquet, M., Carton, J., Cohn, S., Dance, S.,
  Losa, S., Nichols, N., Potthast, R., Waller, J., et~al.: On the
  representation error in data assimilation.
\newblock Quarterly Journal of the Royal Meteorological Society
  \textbf{144}(713), 1257--1278 (2018)

\bibitem{kadakia2016nonlinear}
Kadakia, N., Armstrong, E., Breen, D., Morone, U., Daou, A., Margoliash, D.,
  Abarbanel, H.D.: Nonlinear statistical data assimilation for hvc$\_{RA}$
  neurons in the avian song system.
\newblock Biological cybernetics \textbf{110}(6), 417--434 (2016)

\bibitem{Kuptsov2012}
Kuptsov, P.V., Parlitz, U.: Theory and computation of covariant {L}yapunov
  vectors.
\newblock J. Nonlinear Sci. \textbf{22}, 727--762 (2012)

\bibitem{laloyaux2016coupled}
Laloyaux, P., Balmaseda, M., Dee, D., Mogensen, K., Janssen, P.: A coupled data
  assimilation system for climate reanalysis.
\newblock Q J Roy. Meteor. Soc. \textbf{142}(694), 65--78 (2016)

\bibitem{laloyaux2018cera}
Laloyaux, P., de~Boisseson, E., Balmaseda, M., Bidlot, J.R., Broennimann, S.,
  Buizza, R., Dalhgren, P., Dee, D., Haimberger, L., Hersbach, H., et~al.:
  Cera-20c: A coupled reanalysis of the twentieth century.
\newblock Journal of Advances in Modeling Earth Systems \textbf{10}(5),
  1172--1195 (2018)

\bibitem{Legras1995}
Legras, B., Vautard, R.: A guide to {L}iapunov vectors.
\newblock In: Proceedings 1995 ECMWF Seminar on Predictability, vol.~1, pp.
  143--156. Citeseer (1996)

\bibitem{lorenc20074d}
Lorenc, A.C., Payne, T.: {4D-Var} and the butterfly effect: Statistical
  four-dimensional data assimilation for a wide range of scales.
\newblock Q J Roy. Meteor. Soc. \textbf{133}(624), 607--614 (2007)

\bibitem{lu2015strongly}
Lu, F., Liu, Z., Zhang, S., Liu, Y.: {Strongly coupled data assimilation using
  leading averaged coupled covariance (LACC). Part I: Simple model study}.
\newblock Mon. Weather Rev. \textbf{143}(9), 3823--3837 (2015)

\bibitem{lucarini2020new}
Lucarini, V., Gritsun, A.: A new mathematical framework for atmospheric
  blocking events.
\newblock Climate Dynamics \textbf{54}(1-2), 575--598 (2020)

\bibitem{moye2018data}
Moye, M.J., Diekman, C.O.: Data assimilation methods for neuronal state and
  parameter estimation.
\newblock The Journal of Mathematical Neuroscience \textbf{8}(1), 11 (2018)

\bibitem{ott2002chaos}
Ott, E.: Chaos in dynamical systems.
\newblock Cambridge university press (2002)

\bibitem{palatella2013a}
Palatella, L., Carrassi, A., Trevisan, A.: Lyapunov vectors and assimilation in
  the unstable subspace: theory and applications.
\newblock J. Phys. A: Math. Theor. \textbf{46}, 254020 (2013)

\bibitem{palatella2013nonlinear}
Palatella, L., Trevisan, A., Rambaldi, S.: Nonlinear stability of traffic
  models and the use of lyapunov vectors for estimating the traffic state.
\newblock Phys. Rev. E. \textbf{88}(2), 022901 (2013)

\bibitem{palmer2008toward}
Palmer, T., Doblas-Reyes, F., Weisheimer, A., Rodwell, M.: Toward seamless
  prediction: Calibration of climate change projections using seasonal
  forecasts.
\newblock Bulletin of the American Meteorological Society \textbf{89}(4),
  459--470 (2008)

\bibitem{park2015structure}
Park, S.K., Lim, S., Zupanski, M.: Structure of forecast error covariance in
  coupled atmosphere--chemistry data assimilation.
\newblock Geoscientific Model Development \textbf{8}(5), 1315--1320 (2015)

\bibitem{Pazo2010}
Paz{\'o}, D., Rodr{\'\i}guez, M.A., L{\'o}pez, J.M.: Spatio-temporal evolution
  of perturbations in ensembles initialized by bred, lyapunov and singular
  vectors.
\newblock Tellus A: Dynamic Meteorology and Oceanography \textbf{62}(1), 10--23
  (2010)

\bibitem{penny2017coupledWMO}
Penny, S.G., Akella, S., Buehner, M., Chevallier, M., Counillon, F., Draper,
  C., Frolov, S., Fujii, Y., Karspeck, A., Kumar, A., Laloyaux, P., Mahfouf,
  J.F., Matthew, M., Pe\~na, M., de~Rosnay, P., Subramanian, A., Tardiff, R.,
  Wang, Y., Wu, X.: Coupled data assimilation for integrated earth system
  analysis and prediction: Goals, challenges, and recommendations.
\newblock In: Techincal Report WWRP 2017-3, vol.~3, pp. 1--59. World
  Meteorological Organization (2017)

\bibitem{Penny-2019}
Penny, S.G., Bach, E., Bhargava, K., Chang, C.C., Da, C., Sun, L., Yoshida, T.:
  Strongly coupled data assimilation in multiscale media: Experiments using a
  quasi-geostrophic coupled model.
\newblock Journal of Advances in Modeling Earth Systems \textbf{11}(6),
  1803--1829 (2019).
\newblock \doi{10.1029/2019MS001652}.
\newblock
  \urlprefix\url{https://agupubs.onlinelibrary.wiley.com/doi/abs/10.1029/2019MS001652}

\bibitem{penny2017coupled}
Penny, S.G., Hamill, T.M.: Coupled data assimilation for integrated earth
  system analysis and prediction.
\newblock Bull. Amer. Meteor. Soc. \textbf{97}(7), ES169--ES172 (2017)

\bibitem{npg-27-51-2020}
Quinn, C., O'Kane, T.J., Kitsios, V.: Application of a local attractor
  dimension to reduced space strongly coupled data assimilation for chaotic
  multiscale systems.
\newblock Nonlinear Processes in Geophysics \textbf{27}(1), 51--74 (2020).
\newblock \doi{10.5194/npg-27-51-2020}.
\newblock \urlprefix\url{https://www.nonlin-processes-geophys.net/27/51/2020/}

\bibitem{Ruelle1979}
Ruelle, D.: Ergodic theory of differentiable dynamical systems.
\newblock Publications Math{\'e}matiques de l'Institut des Hautes {\'E}tudes
  Scientifiques \textbf{50}(1), 27--58 (1979)

\bibitem{saha2010ncep}
Saha, S., Moorthi, S., Pan, H.L., Wu, X., Wang, J., Nadiga, S., Tripp, P.,
  Kistler, R., Woollen, J., Behringer, D., et~al.: The ncep climate forecast
  system reanalysis.
\newblock Bull. Amer. Meteor. Soc. \textbf{91}(8), 1015--1057 (2010)

\bibitem{sakov2012topaz4}
Sakov, P., Counillon, F., Bertino, L., Lis{\ae}ter, K., Oke, P., Korablev, A.:
  {TOPAZ4: an ocean-sea ice data assimilation system for the North Atlantic and
  Arctic}.
\newblock Ocean Sci. \textbf{8}(4), 633 (2012)

\bibitem{schepers2018cera}
Schepers, D., de~Boiss{\'e}son, E., Eresmaa, R., Lupu, C., de~Rosnay, P.:
  Cera-sat: A coupled satellite-era reanalysis.
\newblock ECMWF Newslett \textbf{155}, 32--37 (2018)

\bibitem{sluka2016assimilating}
Sluka, T.C., Penny, S.G., Kalnay, E., Miyoshi, T.: Assimilating atmospheric
  observations into the ocean using strongly coupled ensemble data
  assimilation.
\newblock Geophys. Res. Let. \textbf{43}(2), 752--759 (2016)

\bibitem{smith2015exploring}
Smith, P.J., Fowler, A.M., Lawless, A.S.: Exploring strategies for coupled
  {4D-Var} data assimilation using an idealised atmosphere--ocean model.
\newblock Tellus A \textbf{67}(1), 27025 (2015)

\bibitem{smith2017estimating}
Smith, P.J., Lawless, A.S., Nichols, N.K.: Estimating forecast error
  covariances for strongly coupled atmosphere--ocean 4d-var data assimilation.
\newblock Monthly Weather Review \textbf{145}(10), 4011--4035 (2017)

\bibitem{smith2018treating}
Smith, P.J., Lawless, A.S., Nichols, N.K.: Treating sample covariances for use
  in strongly coupled atmosphere-ocean data assimilation.
\newblock Geophysical Research Letters \textbf{45}(1), 445--454 (2018)

\bibitem{sugiura2008development}
Sugiura, N., Awaji, T., Masuda, S., Mochizuki, T., Toyoda, T., Miyama, T.,
  Igarashi, H., Ishikawa, Y.: Development of a four-dimensional variational
  coupled data assimilation system for enhanced analysis and prediction of
  seasonal to interannual climate variations.
\newblock J. Geophys. Res. Oceans \textbf{113}(C10) (2008)

\bibitem{suzuki2017case}
Suzuki, K., Zupanski, M., Zupanski, D.: A case study involving single
  observation experiments performed over snowy siberia using a coupled
  atmosphere-land modelling system.
\newblock Atmospheric Science Letters \textbf{18}(3), 106--111 (2017)

\bibitem{tardif2014coupled}
Tardif, R., Hakim, G.J., Snyder, C.: Coupled atmosphere--ocean data
  assimilation experiments with a low-order climate model.
\newblock Clim. Dyn. \textbf{43}(5-6), 1631--1643 (2014)

\bibitem{tardif2015coupled}
Tardif, R., Hakim, G.J., Snyder, C.: Coupled atmosphere--ocean data
  assimilation experiments with a low-order model and {CMIP5} model data.
\newblock Clim. Dyn. \textbf{45}(5-6), 1415--1427 (2015)

\bibitem{Trevisan1998}
Trevisan, A., Pancotti, F.: Periodic orbits, lyapunov vectors, and singular
  vectors in the lorenz system.
\newblock J. Atmos. Sci. \textbf{55}(3), 390--398 (1998)

\bibitem{vannitsem2015role}
Vannitsem, S.: The role of the ocean mixed layer on the development of the
  north atlantic oscillation: A dynamical system's perspective.
\newblock Geophysical Research Letters \textbf{42}(20), 8615--8623 (2015)

\bibitem{VANNITSEM201571}
Vannitsem, S., Demaeyer, J., Cruz, L.D., Ghil, M.: Low-frequency variability
  and heat transport in a low-order nonlinear coupled ocean–atmosphere model.
\newblock Physica D: Nonlinear Phenomena \textbf{309}, 71 -- 85 (2015).
\newblock \doi{https://doi.org/10.1016/j.physd.2015.07.006}.
\newblock
  \urlprefix\url{http://www.sciencedirect.com/science/article/pii/S0167278915001335}

\bibitem{vannitsem2016statistical}
Vannitsem, S., Lucarini, V.: Statistical and dynamical properties of covariant
  lyapunov vectors in a coupled atmosphere-ocean model—multiscale effects,
  geometric degeneracy, and error dynamics.
\newblock Journal of Physics A: Mathematical and Theoretical \textbf{49}(22),
  224001 (2016)

\bibitem{Wolfe2007}
Wolfe, C.L., Samelson, R.M.: An efficient method for recovering lyapunov
  vectors from singular vectors.
\newblock Tellus A: Dynamic Meteorology and Oceanography \textbf{59}(3),
  355--366 (2007)

\bibitem{zhang2007system}
Zhang, S., Harrison, M., Rosati, A., Wittenberg, A.: System design and
  evaluation of coupled ensemble data assimilation for global oceanic climate
  studies.
\newblock Mon. Weather Rev. \textbf{135}(10), 3541--3564 (2007)

\bibitem{zupanski2005}
Zupanski, M.: Maximum likelihood ensemble filter: {T}heoretical aspects.
\newblock Mon. Weather Rev. \textbf{133}, 1710--1726 (2005)

\bibitem{vzupanski2017data}
Zupanski, M.: Data assimilation for coupled modeling systems.
\newblock In: Data Assimilation for Atmospheric, Oceanic and Hydrologic
  Applications (Vol. III), pp. 55--70. Springer (2017)

\bibitem{acp-2019-2}
Zupanski, M., Kliewer, A., Wu, T.C., Apodaca, K., Bian, Q., Atwood, S., Wang,
  Y., Wang, J., Miller, S.D.: Impact of atmospheric and aerosol optical depth
  observations on aerosol initial conditions in a strongly-coupled data
  assimilation system.
\newblock Atmospheric Chemistry and Physics Discussions \textbf{2019}, 1--25
  (2019).
\newblock \doi{10.5194/acp-2019-2}.
\newblock \urlprefix\url{https://www.atmos-chem-phys-discuss.net/acp-2019-2/}

\end{thebibliography}

\end{document}